\documentclass[notitlepage,nofootinbib,preprintnumbers,amssymb,superscriptaddress]{revtex4-1}
\usepackage{amsfonts,amssymb,amsmath,graphicx,color,bm}
\definecolor{ultramarine}{rgb}{0.07, 0.04, 0.56}
\definecolor{cadmiumgreen}{rgb}{0.0, 0.42, 0.24}
\definecolor{indigo(dye)}{rgb}{0.0, 0.25, 0.42}
\usepackage[normalem]{ulem}
\usepackage[linktocpage=true]{hyperref}
\hypersetup{
colorlinks=true,
citecolor=green,
linkcolor=cadmiumgreen,
urlcolor=indigo(dye),
}
\usepackage{eufrak}
\usepackage[normalem]{ulem}
\usepackage{graphicx,color,epsfig}
\usepackage{wrapfig}

\begin{document}

\title{Monopole stability of time-dependent, spherically symmetric backgrounds in Horndeski theory}

\author{K. Kiselev}
\email{kiselev.ka21@physics.msu.ru}
\affiliation{Institute for Nuclear Research of the Russian Academy of Sciences,
60th October Anniversary Prospect, 7a, 117312 Moscow, Russia}
\affiliation{Department of Particle Physics and Cosmology, Physics Faculty, M.V. Lomonosov Moscow State University,
Vorobjevy Gory, 119991 Moscow, Russia}

\author{V. Volkova}
\email{volkova.viktoriya@physics.msu.ru}
\affiliation{Institute for Nuclear Research of the Russian Academy of Sciences,
60th October Anniversary Prospect, 7a, 117312 Moscow, Russia}
\affiliation{Department of Particle Physics and Cosmology, Physics Faculty, M.V. Lomonosov Moscow State University,
Vorobjevy Gory, 119991 Moscow, Russia}

\begin{abstract}
We study linear perturbations about an arbitrary time-dependent, spherically symmetric background in the most general Horndeski theory. The odd-parity sector of such perturbations, analyzed previously, contains no scalar field perturbation and, hence, leaves the related degree of freedom unconstrained.
In this paper, we address this issue by analyzing even-parity monopole ($\ell = 0$) perturbations, which isolate this degree of freedom. 
We derive the quadratic master action for the only propagating physical mode associated with the scalar field and formulate stability criteria which ensure the absence of ghost and radial gradient instabilities for the high-energy modes.
These conditions are necessary for the stability of the background and complement the odd-parity ones. We show that they reproduce the known results in the static and cosmological limits. 
We also demonstrate that, in the cubic subclass of Horndeski theory on a dynamical background, the monopole criteria impose no constraints beyond those previously obtained for modes with large multipoles ($\ell \gg 1$).
As an illustration, we apply the derived stability conditions for the monopole mode to a non-minimally coupled scalar field theory and to a black hole model with linearly time-dependent scalar hair; in the latter case, the monopole conditions further restrict the parameter space allowed by odd-parity stability.
\end{abstract}

\maketitle
   
    \section{Introduction}
    \label{sec:intro}

Scalar-tensor theories have become one of the most extensively studied frameworks for modifying General Relativity (GR), incorporating an additional scalar field. 
The most general four-dimensional scalar-tensor theory with second-order equations of motion was formulated by Horndeski~\cite{Horndeski1974} and later rediscovered in the form of generalized Galileons~\cite{Nicolis2009,Deffayet2009generalized,Deffayet2011,Kobayashi2011_flrw}, which inherently evade the Ostrogradsky ghost. 
Second-order equations of motion are, however, sufficient but not necessary for the absence of the Ostrogradsky ghost: in beyond Horndeski (GLPV) theories~\cite{Zumalacarregui2014,Gleyzes2015,Gleyzes2015exploring} and, more generally, in degenerate higher-order scalar-tensor (DHOST) theories~\cite{LangloisNoui2016,Crisostomi2016,BenAchour2016disformal,BenAchour2016cubic}, the equations of motion are of higher order, but the degeneracy of the kinetic matrix eliminates the would-be extra mode. These classes of theories constitute the most general scalar-tensor gravity, which propagates a single scalar degree of freedom in addition to the two polarizations of the graviton (see Refs.~\cite{Kobayashi2019review,Langlois2019review} for reviews).

The application scope of Horndeski theory and its generalizations is exceptionally broad. In a cosmological setting they support early Universe models like inflation 
and, owing to the possibility of a healthy violation of the null energy condition (NEC)~\cite{Rubakov2014}, non-singular Genesis~\cite{Creminelli2010genesis} and bouncing~\cite{Easson2011} scenarios (see Refs.~\cite{Heisenberg:2018vsk,MironovVolkova2024review} for a review). At late times, Horndeski theories can describe dynamical dark energy~\cite{Deffayet2010KGB,DeFeliceTsujikawa2012,Gleyzes2013,KaseTsujikawa2019} and, despite the stringent constraints imposed by the multi-messenger speed test GW170817~\cite{LIGO2017GW170817,Ezquiaga2017,CreminelliVernizzi2017,Baker2017,Langlois2018}, this direction has attracted renewed interest in light of the recent DESI results~\cite{DESI2025DR2,Ye2025}. 
On astrophysical scales, Horndeski theory and its generalizations admit hairy and spontaneously scalarized black holes and neutron stars (see e.g. Refs.~\cite{Babichev2016review,Babichev2023book,BenAchour2025review} for reviews), as well as traversable wormholes supported by the NEC violation, see e.g. Refs.~\cite{Mironov2019wormhole,Mironov2023wormhole}.  
The class of black hole solutions with time-dependent scalar hair is of particular interest: in shift-symmetric theories, the scalar field profile $\pi = qt + \psi(r)$ is compatible with a static, spherically symmetric metric~\cite{BabichevCharmousis2014}, which makes such configurations possibly the simplest example of a background with a genuinely time-dependent scalar field.

The existence of a background solution is, however, only the first step towards establishing its viability: the solution has to be stable, at least at the linearized level. Conditions ensuring the absence of ghosts and gradient instabilities have been derived in Horndeski theory and its generalizations for different backgrounds, including cosmological ones~\cite{Kobayashi2011_flrw,DeFeliceTsujikawa2012,Gleyzes2013,Kolevatov2017perturbations,MironovShtennikova2023}, static spherically symmetric ones~\cite{Kobayashi2012_static,Kobayashi2014_static,Minamitsuji2022b,MironovVolkova2023DPSV,MironovVolkova2024complete,Boonserm2026} such as wormholes~\cite{Mironov2019wormhole,Mironov2023wormhole} and neutron stars~\cite{BoumazaLanglois2025}. 
The linear stability of black holes with scalar hair in shift-symmetric theories, including those with a linearly time-dependent scalar field, has been studied mainly in the axial (odd-parity) sector~\cite{Ogawa2016,TakahashiSuyama2017,TretyakovaTakahashi2017,Babichev2018PRL,Takahashi2019,TomikawaKobayashi2021,Noui2023,Charmousis2025axial}, while the polar (even-parity) sector, which contains the perturbation associated with the scalar field, has been addressed in fewer works~\cite{Babichev2018hamiltonian,deRhamZhang2019,Langlois2021,TakahashiMotohashi2021,Minamitsuji2022a,Charmousis2026radial}.
In several cases, the stability conditions turned out to be incompatible with a healthy and complete solution, which resulted in no-go theorems for non-singular cosmologies~\cite{Libanov2016,Kobayashi2016nogo} and static wormholes~\cite{Rubakov2016wormholes,EvseevMelichev2018}.

 Recently, linear stability on an arbitrary dynamical, spherically symmetric background, where all metric functions and the scalar field are time-dependent, was examined in Ref.~\cite{oddSector}. The analysis was restricted to the odd-parity sector of perturbations, while corresponding high-energy stability constraints were formulated for a subclass of GLPV theories with a non-trivial $F_4$ scalar potential. 
 However, the odd-parity sector contains only one of the graviton polarizations, while the scalar field does not fluctuate at the linear level. Hence, the odd-parity analysis leaves the scalar field perturbations unconstrained. So far the even-parity sector on a general dynamical, spherically symmetric background has been analyzed only in the cubic subclass of Horndeski theory~\cite{MironovSharov2025}, where the gravitational sector coincides with that of GR, and the resulting stability conditions were used to examine whether the static and cosmological no-go theorems extend to dynamical, spherically symmetric backgrounds. The method employed in Ref.~\cite{MironovSharov2025}, however, does not straightforwardly extend beyond the cubic subclass on spherically symmetric backgrounds~\cite{MironovVolkova2023DPSV}. Thus, the even-parity sector of the general Horndeski theory on dynamical, spherically symmetric backgrounds so far remains unexplored.

 In this work we take the first step towards closing this gap and study the even-parity monopole ($\ell = 0$) perturbations about an arbitrary time-dependent, spherically symmetric background in full Horndeski theory. 
 The main reason for focusing on the monopole sector is that it isolates the single physical degree of freedom associated with the scalar field.
 Generally, the low-multipole modes, and specifically the monopole, require a separate treatment: for $\ell = 0$ all angular derivatives vanish identically, the gauge freedom can be fixed completely and all auxiliary metric perturbations can be integrated out explicitly even for a fully dynamical, spherically symmetric background, where calculations become significantly more complicated. 
 This allows us to derive the quadratic action for the master variable and formulate the conditions for the absence of ghosts and radial gradient instabilities for high-energy modes as well as derive the corresponding radial propagation speeds. We do not consider slow tachyonic instabilities, which are governed by the effective potential in the resulting action. We stress that, by construction, the monopole analysis is insensitive to perturbations propagating in the angular directions, which are known to be a source of instabilities in a number of cases, see e.g. Refs.~\cite{Tsujikawa2022,Minamitsuji2024}. Moreover, for dynamical backgrounds an independent angular stability condition arises already in the cubic Horndeski subclass~\cite{MironovSharov2025}. 
 Hence, the stability conditions derived in this paper are necessary, but not sufficient, for the stability of the background. The analysis of the dipole ($\ell = 1$) and higher multipoles on a fully dynamical background is considerably more involved, since the elimination of non-dynamical variables and the derivation of the master equations become technically demanding; we leave it for future work.

 This paper is organized as follows. In Sec.~\ref{sec:HorndeskiTheory} we introduce Horndeski theory and the dynamical, spherically symmetric background. In Sec.~\ref{sec:odd_review}, we briefly review the results for the odd-parity stability conditions derived in Ref.~\cite{oddSector}. In Sec.~\ref{sec:even_perturbations} we set up the even-parity perturbations, and in Sec.~\ref{sec:monopole} we fix the gauge, integrate out the non-dynamical metric perturbations and derive the quadratic action for the monopole master variable. The high-energy stability conditions and the radial propagation speeds are obtained in Sec.~\ref{sec:monopole_stability}. In Sec.~\ref{sec:limiting_cases}, we show that, in the corresponding limits, our results reproduce the known propagation speeds for static, spherically symmetric~\cite{Kobayashi2014_static} and spatially flat FLRW~\cite{Kobayashi2011_flrw} backgrounds. We also show there that, in the cubic subclass of Horndeski theory on a dynamical, spherically symmetric background, the monopole stability conditions impose no additional constraints beyond those obtained for high-momentum modes in Ref.~\cite{MironovSharov2025}.
 In Sec.~\ref{sec:applications}, we apply our stability criteria to a non-minimally coupled scalar field and to a black hole solution with linearly time-dependent scalar hair. We conclude in Sec.~\ref{sec:conclusions}. The background equations of motion and the coefficients of the quadratic action are collected in Appendices~\ref{app:background_eom}--\ref{app:auxiliary_shift}.

\section{Dynamical spherically symmetric background in Horndeski theory}
    \label{sec:HorndeskiTheory}
        We consider Horndeski theory described by the four-dimensional action~\cite{Horndeski1974,Kobayashi2019review}
        \begin{equation}
            S = \int \mathrm{d}^{4}x \sqrt{-g} \left( \mathcal{L}_{2} + \mathcal{L}_{3} + \mathcal{L}_{4} + \mathcal{L}_{5} \right),
            \label{eq:action}
        \end{equation}
        where the individual Lagrangian densities are given by
        \begin{subequations}
        \label{eq:Lagrangians}
        \begin{align}
            \mathcal{L}_{2} &= F(\pi,X), \label{eq:L2} \\
            \mathcal{L}_{3} &= -K(\pi,X)\Box\pi, \label{eq:L3} \\
            \mathcal{L}_{4} &= G_{4}(\pi,X)R + G_{4X}(\pi,X)\left[(\Box\pi)^{2} - \pi_{;\mu\nu}\pi^{;\mu\nu}\right], \label{eq:L4} \\
            \mathcal{L}_{5} &= G_{5}(\pi,X)G^{\mu\nu}\pi_{;\mu\nu} - \frac{1}{6}G_{5X}(\pi,X)\left[(\Box\pi)^{3} - 3\Box\pi\pi_{;\mu\nu}\pi^{;\mu\nu} + 2\pi_{;\mu\nu}\pi^{;\mu\rho}\pi^{\nu}_{\phantom{\nu};\rho}\right]. \label{eq:L5}
        \end{align}
        \end{subequations}
        Here, $\pi$ is the scalar field, $R$ is the Ricci scalar, $G_{\mu\nu}$ is the Einstein tensor, and $F$, $K$, $G_{4}$, and $G_{5}$ are arbitrary functions of $\pi$ and the canonical kinetic term
        \begin{equation}
            X = -\frac{1}{2} g^{\mu\nu}\pi_{,\mu}\pi_{,\nu}.
            \label{eq:kinetic_term}
        \end{equation}
        We adopt the shorthand notation $\pi_{,\mu} \equiv \partial_{\mu}\pi$, $\pi_{;\mu\nu} \equiv \nabla_{\nu}\nabla_{\mu}\pi$, $\Box\pi \equiv g^{\mu\nu}\nabla_{\nu}\nabla_{\mu}\pi$, and $G_{iX} \equiv \partial G_{i}/\partial X$.
        
        We study linear perturbations about a general time-dependent, spherically symmetric background described by the line element
        \begin{equation}
            \mathrm{d}s^{2} = -A(t,r)\,\mathrm{d}t^{2} + \frac{\mathrm{d}r^{2}}{B(t,r)} + J^{2}(t,r)\gamma_{ab}\,\mathrm{d}x^{a}\mathrm{d}x^{b},
            \label{eq:bg_metric}
        \end{equation}
        where $\gamma_{ab} = \operatorname{diag}(1, \sin^{2}\theta)$ is the standard metric on a unit two-sphere with $a,b \in \{\theta, \phi\}$. The metric functions $A(t,r)$, $B(t,r)$, and $J(t,r)$, as well as the background scalar field profile $\pi = \pi(t,r)$, are assumed to be arbitrary smooth functions of time $t$ and the radial coordinate $r$. Depending on the specific solution to the background field equations, the Ansatz~\eqref{eq:bg_metric} accommodates a wide range of physically relevant configurations, including dynamical black holes, gravitational collapse, pulsating compact objects, etc.

        In what follows, we investigate the behavior of linear perturbations propagating on a dynamical background~\eqref{eq:bg_metric}. The explicit form of the background equations of motion for the metric and the scalar field in theory~\eqref{eq:Lagrangians}, denoted by $\mathcal{E}^{(\mu\nu)} = 0$ and $\mathcal{E}^{(\pi)} = 0$, is given in Appendix~\ref{app:background_eom}.

 \section{Linearized theory}   
     We consider linear metric perturbations $h_{\mu\nu}$ on the dynamical background metric $\bar{g}_{\mu\nu}$~\eqref{eq:bg_metric}:
    \begin{equation}
        g_{\mu\nu} = \bar{g}_{\mu\nu} + h_{\mu\nu}.
        \label{eq:metric_split}
    \end{equation}
    Following the Regge-Wheeler formalism~\cite{ReggeWheeler1957}, metric perturbations are classified into odd-parity and even-parity sectors according to their transformation properties under reflections on the two-sphere. Expanding the perturbation variables in spherical harmonics $Y_{\ell m}(\theta,\phi)$, the background spherical symmetry ensures that modes with different parity, $\ell$, and $m$ decouple at the linear level. Consequently, each sector can be investigated independently.

\subsection{Odd-parity perturbations: a brief review}
\label{sec:odd_review}

    Before analyzing the even-parity sector, we briefly review the odd-parity perturbations over the same dynamical, spherically symmetric background~\eqref{eq:bg_metric}. Recently, a linear stability analysis for this sector was performed in Ref.~\cite{oddSector} for beyond Horndeski theories. Here, we present the results specialized to the Horndeski limit ($F_4 = 0$).
    
    In the odd-parity sector, the scalar field remains unperturbed ($\delta\pi_{\text{odd}} = 0$). The purely gravitational dynamics reduce to a single propagating mode, associated with the one of the graviton polarizations. Its high-energy behavior is governed by four background-dependent coefficients\footnote{In this work the coefficient $\mathcal{D}$ corresponds to $\mathcal{J}$ in Ref.~\cite{oddSector}.}:
    \begin{subequations}
    \label{eq:odd_coeffs}
    \begin{align}
        \mathcal{F} &= 2 \left[ G_4 + X G_{5X} \left( \frac{\dot{\pi}}{2A} \frac{\dot{B}}{B} - \frac{1}{\pi'} \frac{\mathrm{d}}{\mathrm{d}r} \left( -\frac{1}{2} B \pi'^2 \right) \right) - G_{5\pi} \left( X - \frac{\dot{\pi}^2}{A} \right) - \frac{\dot{\pi}^2}{A} G_{4X} \right], \label{eq:odd_F} \\
        \mathcal{G} &= 2 \left[ G_4 - 2 G_{4X} \left( X - \frac{\dot{\pi}^2}{2A} \right) + G_{5\pi} \left( X - \frac{\dot{\pi}^2}{A} \right) + X G_{5X} \left( \frac{A'}{2A} B \pi' + \frac{\dot{A} \dot{\pi}}{2 A^2} - \frac{\ddot{\pi}}{A} \right) \right], \label{eq:odd_G} \\
        \mathcal{H} &= 2 \left[ G_4 - 2X G_{4X} + X \left( \left( \frac{B J' \pi'}{J} - \frac{\dot{J}\dot{\pi}}{A J} \right) G_{5X} + G_{5\pi} \right) \right]=-\frac{2A}{B}\frac{\partial \mathcal{E}^{(22)}}{\partial A''}, \label{eq:odd_H} \\
        \mathcal{D} &= 2\dot{\pi}\pi' (G_{4X} - G_{5\pi}) - X G_{5X} \left( 2\dot{\pi}' - \frac{A'\dot{\pi}}{A} + \frac{\dot{B}\pi'}{B} \right), \label{eq:odd_J}
    \end{align}
    \end{subequations}
    where $\pi$ refers to the background scalar field profile, while dots and primes denote derivatives with respect to $t$ and $r$, respectively.
    
    The absence of ghost and gradient instabilities requires the kinetic term of the sole odd-parity dynamical mode to be positive and the propagation speeds to be real. This imposes the following simultaneous stability conditions:
    \begin{equation}
        \mathcal{F} > 0, \qquad \mathcal{Z} > 0, \qquad \mathcal{H} > 0,
        \label{eq:odd_stability}
    \end{equation}
    where the auxiliary variable is defined as $\mathcal{Z} \equiv \mathcal{G}\mathcal{F} + \dfrac{B}{A} \mathcal{D}^2$. 
    
    The radial and angular propagation speeds are given by, respectively:
    \begin{equation}
        c_{r,\, \text{odd}}^{(\pm)} = \sqrt{\frac{B}{A}} \frac{\mathcal{D}}{\mathcal{F}} \pm \frac{1}{\mathcal{F}} \sqrt{\mathcal{Z}}, \qquad c_{\theta,\, \text{odd}}^2 = \frac{\mathcal{Z}}{\mathcal{F}\mathcal{H}}.
        \label{eq:odd_speeds}
    \end{equation}
    
    As shown below, the coefficients $\mathcal{F}$, $\mathcal{G}$, $\mathcal{D}$, and $\mathcal{H}$ also constitute the fundamental building blocks for the even-parity stability criteria.

\subsection{Even-parity perturbations on a time-dependent, spherically symmetric background}\label{sec:even_perturbations}
    
    Let us consider the even-parity sector with the following parametrization of metric perturbations:
    \begin{subequations}
    \label{eq:even_metric_pert}
    \begin{align}
        h_{tt} &= A(t,r)\sum_{\ell,m} H_{0,\ell m}(t,r) Y_{\ell m}(\theta,\phi), \label{eq:htt} \\
        h_{tr} &= \sum_{\ell,m} H_{1,\ell m}(t,r) Y_{\ell m}(\theta,\phi), \label{eq:htr} \\
        h_{rr} &= \frac{1}{B(t,r)}\sum_{\ell,m} H_{2,\ell m}(t,r) Y_{\ell m}(\theta,\phi), \label{eq:hrr} \\
        h_{ta} &= \sum_{\ell,m} \beta_{\ell m}(t,r) \partial_{a} Y_{\ell m}(\theta,\phi), \label{eq:hta} \\
        h_{ra} &= \sum_{\ell,m} \alpha_{\ell m}(t,r) \partial_{a} Y_{\ell m}(\theta,\phi), \label{eq:hra} \\
        h_{ab} &= \sum_{\ell,m} \left[ K_{\ell m}(t,r) \bar{g}_{ab} Y_{\ell m}(\theta,\phi) + G_{\ell m}(t,r) \nabla_{a}\nabla_{b}Y_{\ell m}(\theta,\phi) \right], \label{eq:hab}
    \end{align}
    \end{subequations}
    where $\bar{g}_{ab} = J^2(t,r)\gamma_{ab}$, and $\nabla_a$ denotes the covariant derivative compatible with the metric $\gamma_{ab}$ on the unit two-sphere, with indices $a,b \in \{\theta, \phi\}$. In contrast to the odd-parity sector, the scalar field $\pi$ also acquires a non-trivial perturbation:
    \begin{equation}
        \pi(t,r,\theta,\phi) = \pi(t,r) + \sum_{\ell,m}\chi_{\ell m}(t,r) Y_{\ell m}(\theta,\phi),
        \label{eq:scalar_pert}
    \end{equation}
    where $\pi(t,r)$ is the background scalar field. 
    
    General covariance generates gauge redundancy, which allows one to eliminate a subset of the perturbation variables through an infinitesimal coordinate transformation $x^{\mu} \to x^{\mu} + \xi^{\mu}$. Three of the four gauge degrees of freedom belong to the even-parity sector. The gauge vector $\xi_\mu$ can be decomposed into spherical harmonics as
   \begin{equation}
    \xi_{t} = \sum_{\ell,m} T_{\ell m}(t,r) Y_{\ell m}(\theta,\phi), \quad
    \xi_{r} = \sum_{\ell,m} R_{\ell m}(t,r) Y_{\ell m}(\theta,\phi), \quad
    \xi_{a} = \sum_{\ell,m} \Theta_{\ell m}(t,r) \partial_{a} Y_{\ell m}(\theta,\phi),
    \label{eq:gauge_vector}
    \end{equation}
    where $T_{\ell m}(t,r)$, $R_{\ell m}(t,r)$, and $\Theta_{\ell m}(t,r)$ are arbitrary gauge-generating functions.
    
    Under these infinitesimal coordinate transformations, the metric and scalar field perturbations transform as follows:
    \begin{align}
        h_{\mu\nu} &\to h_{\mu\nu} + \nabla_\mu \xi_\nu + \nabla_\nu \xi_\mu , \label{eq:gauge_h_general} \\
        \delta\pi &\to \delta\pi + \xi^{\mu}\partial_{\mu}\pi. \label{eq:gauge_pi_general}
    \end{align}
    Hence, for a dynamical background~\eqref{eq:bg_metric} the gauge transformation laws for perturbations variables in Eqs.~\eqref{eq:even_metric_pert} explicitly read:
    \begin{subequations}
    \label{eq:even_gauge_transformations}
    \begin{align}
        H_{0,\ell m} &\to H_{0,\ell m} + \frac{2\dot{T}_{\ell m}}{A} - \frac{A' B}{A} R_{\ell m} - \frac{\dot{A}}{A^2} T_{\ell m}, \label{eq:transf_H0} \\
        H_{1,\ell m} &\to H_{1,\ell m} + \dot{R}_{\ell m} + T'_{\ell m} - \frac{A'}{A} T_{\ell m} - \frac{\dot{B}}{B} R_{\ell m}, \label{eq:transf_H1} \\
        H_{2,\ell m} &\to H_{2,\ell m} + 2 B R'_{\ell m} + B' R_{\ell m} + \frac{\dot{B}}{A B} T_{\ell m}, \label{eq:transf_H2} \\
        \beta_{\ell m} &\to \beta_{\ell m} + T_{\ell m} + \dot{\Theta}_{\ell m} - \frac{2\dot{J}}{J}\Theta_{\ell m}, \label{eq:transf_beta} \\
        \alpha_{\ell m} &\to \alpha_{\ell m} + R_{\ell m} + \Theta'_{\ell m} - \frac{2J'}{J}\Theta_{\ell m}, \label{eq:transf_alpha} \\
        K_{\ell m} &\to K_{\ell m} + \frac{2B J'}{J} R_{\ell m} - \frac{2\dot{J}}{A J} T_{\ell m}, \label{eq:transf_K} \\
        G_{\ell m} &\to G_{\ell m} + 2\Theta_{\ell m}, \label{eq:transf_G} \\
        \chi_{\ell m} &\to \chi_{\ell m} + B \pi' R_{\ell m} - \frac{\dot{\pi}}{A} T_{\ell m} \label{eq:transf_chi}.
    \end{align}
    \end{subequations}
    In the static limit, i.e. with $A=A(r), B=B(r), J=J(r), \pi=\pi(r)$ in Eq.~\eqref{eq:bg_metric}, the time derivatives of the background functions vanish, so one straightforwardly recovers the standard gauge transformation laws for static backgrounds, see e.g. Ref.~\cite{Kobayashi2014_static}.

\subsection{Monopole perturbations in Horndeski theory: $\ell=0$}\label{sec:monopole}

    We now focus on the monopole mode ($\ell = 0$). For perturbations with $\ell = 0$ the spherical harmonic reduces to a constant, $Y_{00} = 1/\sqrt{4\pi}$, so that all angular derivatives vanish identically and the perturbation variables associated with angular derivatives in Eq.~\eqref{eq:even_metric_pert} are absent from the outset: $\beta = \alpha = G \equiv 0$. The gauge vector~\eqref{eq:gauge_vector} likewise loses its angular component ($\Theta \equiv 0$), leaving only two independent gauge-generating functions, $T(t,r)$ and $R(t,r)$.
    
    According to the transformation laws~\eqref{eq:transf_K} and~\eqref{eq:transf_chi}, requiring both transformed quantities to vanish yields a linear algebraic system for $T(t,r)$ and $R(t,r)$:
    \begin{equation}
        \begin{pmatrix}
            -\dfrac{2\dot{J}}{A J} & \dfrac{2 B J'}{J} \\[8pt]
            -\dfrac{\dot{\pi}}{A} & B \pi'
        \end{pmatrix}
        \begin{pmatrix}
            T \\ R
        \end{pmatrix}
        =
        \begin{pmatrix}
            -K \\ -\chi
        \end{pmatrix},
        \label{eq:gauge_matrix_system}
    \end{equation}
    with a determinant of this system given by
    \begin{equation}
        \Delta = \frac{2 B}{A J} \left( J' \dot{\pi} - \dot{J} \pi' \right).
        \label{eq:gauge_det}
    \end{equation}
    Since $A, B, J > 0$, the system~\eqref{eq:gauge_matrix_system} possesses a unique algebraic solution provided that
    \begin{equation}
        J' \dot{\pi} - \dot{J} \pi' \neq 0.
        \label{eq:non_degeneracy_cond}
    \end{equation}
    For static backgrounds, one has $\dot{\pi} = \dot{J} = 0$, so that the condition~\eqref{eq:non_degeneracy_cond} no longer holds, reflecting the degeneracy of this gauge choice in the static limit. Nevertheless, as we show in Sec.~\ref{sec:limiting_cases} below, the physical radial propagation speeds remain regular and smoothly recover the known static limit upon taking $\dot{\pi},\dot{J} \to 0$.
    
    Under condition~\eqref{eq:non_degeneracy_cond}, we can unambiguously impose the gauge
    \begin{equation}
        K = \chi = 0.
        \label{eq:monopole_gauge}
    \end{equation}
    Then expanding the action~\eqref{eq:action} to the second order in perturbations, integrating over the two-sphere, and imposing the gauge conditions~\eqref{eq:monopole_gauge}, we obtain the quadratic action for the monopole mode:
    \begin{align}
    S^{(2)}_{\ell=0} = \int \mathrm{d}t \, \mathrm{d}r \ \Bigg[ & n_0 H_0^2 + H_0 \left( a_3 H_2' + a_7 H_2 + (n_2-n_{13}') H_1 + n_3 \dot{H}_2 -n_{13}H_1'\right) \nonumber \\
    & + H_1 \left( b_4 \dot{H}_2 + n_9 H_2 \right) + b_1 H_1^2 + c_6 H_2^2 \Bigg].
    \label{eq:monopole_action}
    \end{align}
    The background-dependent coefficients $n_i, a_i, b_i, c_i$ are determined by the Horndeski functions $F, K, G_4, G_5$ and the background metric functions~\eqref{eq:bg_metric}. Their explicit and rather lengthy expressions are given in Appendix~\ref{app:monopole_coeffs}. It is worth noting that all coefficients denoted by $n_i$ identically vanish in the static limit, while the rest of coefficients recover their form derived for a static, spherically symmetric background in Ref.~\cite{Kobayashi2014_static}. 
    
   Varying the action~\eqref{eq:monopole_action} with respect to $H_0$ yields an algebraic constraint that can be solved explicitly for $H_0$:
    \begin{equation}
        H_0 = -\frac{1}{2 n_0} \left[ n_3 \dot{H}_2 + a_3 H_2' + a_7 H_2 + (n_2 - n_{13}') H_1 - n_{13} H_1' \right].
        \label{eq:H0_constraint}
    \end{equation}
    
    Substituting Eq.~\eqref{eq:H0_constraint} back into the action~\eqref{eq:monopole_action} eliminates $H_0$, yielding the action that depends solely on $H_1$ and $H_2$:
    \begin{align}
        S_{\ell=0}^{(2)} = \int \mathrm{d}t \, \mathrm{d}r \, \Bigg[ & -\frac{1}{4 n_0} \left( n_3 \dot{H}_2 + a_3 H_2' + a_7 H_2 + (n_2 - n_{13}') H_1 - n_{13} H_1' \right)^2 \nonumber \\
        & + b_4 H_1 \dot{H}_2 + n_9 H_1 H_2 + b_1 H_1^2 + c_6 H_2^2 \Bigg].
        \label{eq:monopole_action_intermediate}
    \end{align}
    
   In order to integrate out the auxiliary perturbations $H_1$ and $H_2$ algebraically, all of their spacetime derivatives must be collected into a single perfect square. To absorb the remaining cross-derivative term $b_4 H_1 \dot{H}_2$ into this perfect square without generating extraneous derivative terms, we shift the linear combination inside the perfect square by $k_1 H_1 + k_2 H_2$. Adding and subtracting this shift in Eq.~\eqref{eq:monopole_action_intermediate}, and integrating by parts, the action takes the intermediate form:
    \begin{align}
        S_{\ell=0}^{(2)} = \int \mathrm{d}t \, \mathrm{d}r \, \Bigg[ & -\frac{1}{4 n_0} \left( n_3 \dot{H}_2 + a_3 H_2' - n_{13} H_1' + (n_2 - n_{13}' + k_1) H_1 + (a_7 + k_2) H_2 \right)^2 \nonumber \\
        & + \left( b_4 + \frac{k_1 n_3}{2 n_0} \right) H_1 \dot{H}_2 + \left( \frac{a_3 k_1 + k_2 n_{13}}{2 n_0} \right) H_1 H_2'  + \tilde{\mathcal{M}}_1 H_1^2 + \tilde{\mathcal{M}}_{12} H_1 H_2 + \tilde{\mathcal{M}}_2 H_2^2 \Bigg],
        \label{eq:monopole_action_shifted}
    \end{align}
    where we do not present the explicit form of the intermediate coefficients $\tilde{\mathcal{M}}_1, \tilde{\mathcal{M}}_{12}$, and $\tilde{\mathcal{M}}_2$ for brevity.
    To eliminate the cross-derivative terms $H_1 \dot{H}_2$ and $H_1 H_2'$ outside the squared bracket in Eq.~\eqref{eq:monopole_action_shifted}, we choose the shift parameters $k_1$ and $k_2$ such that their prefactors vanish:
    \begin{equation}
        k_1 = -\frac{2 b_4 n_0}{n_3}, \qquad k_2 = \frac{2 b_4 n_0 a_3}{n_3 n_{13}}.
        \label{eq:k_shifts}
    \end{equation}
    The action then simplifies to
    \begin{align}
        S_{\ell=0}^{(2)} = \int \mathrm{d}t \, \mathrm{d}r \, \Bigg[ & -\frac{1}{4 n_0} \left( n_3 \dot{H}_2 + a_3 H_2' - n_{13} H_1' + (n_2 - n_{13}' + k_1) H_1 + (a_7 + k_2) H_2 \right)^2 \nonumber \\
        & + \mathcal{M}_1 H_1^2 + \mathcal{M}_{12} H_1 H_2 + \mathcal{M}_2 H_2^2 \Bigg],
        \label{eq:monopole_action_completed}
    \end{align}
    where the coefficients $\mathcal{M}_1$, $\mathcal{M}_{12}$, and $\mathcal{M}_2$ are explicitly given by:
    \begin{subequations}
    \label{eq:M_coeffs}
    \begin{align}
        \mathcal{M}_1 &= b_1 + \frac{b_4}{n_3}\left( \frac{b_4 n_0}{n_3} - n_2 + \frac{n_{13}'}{2} \right) - \frac{n_{13}}{2}\left(\frac{b_4}{n_3}\right)', \label{eq:M1} \\[6pt]
        \mathcal{M}_{12} &= n_9 - \frac{a_7 b_4}{n_3} - \frac{a_3 b_4}{n_3 n_{13}}\left(\frac{2 b_4 n_0}{n_3} - n_2\right) + n_{13}\left(\frac{a_3 b_4}{n_3 n_{13}}\right)', \label{eq:M12} \\[6pt]
        \mathcal{M}_2 &= c_6 + \frac{a_3 b_4}{n_3 n_{13}}\left( a_7 + \frac{a_3 b_4 n_0}{n_3 n_{13}} \right) - \frac{1}{2}\left(\frac{a_3^2 b_4}{n_3 n_{13}}\right)' - \frac{1}{2}\partial_t\left(\frac{a_3 b_4}{n_{13}}\right). \label{eq:M2}
    \end{align}
    \end{subequations}
    Similarly to the odd-parity sector, we linearize the squared derivative term in Eq.~\eqref{eq:monopole_action_completed} by introducing an auxiliary scalar field $Q(t,r)$:
    \begin{align}
        S_{\ell=0}^{(2)} = \int \mathrm{d}t \, \mathrm{d}r \, \bigg[ & -\frac{1}{4 n_0} \Big\{ 2Q \left( n_3 \dot{H}_2 + a_3 H_2' - n_{13} H_1' + (n_2 - n_{13}' + k_1) H_1 + (a_7 + k_2) H_2 \right) - Q^2 \Big\} \nonumber \\
        & + \mathcal{M}_1 H_1^2 + \mathcal{M}_{12} H_1 H_2 + \mathcal{M}_2 H_2^2 \bigg].
        \label{eq:monopole_action_Q}
    \end{align}
    Varying Eq.~\eqref{eq:monopole_action_Q} with respect to $Q$ yields the on-shell condition $Q = n_3 \dot{H}_2 + a_3 H_2' - n_{13} H_1' + (n_2 - n_{13}' + k_1) H_1 + (a_7 + k_2) H_2$, which identically restores the action~\eqref{eq:monopole_action_completed}.
    Integrating by parts shifts all spacetime derivatives from $H_1$ and $H_2$ onto $Q$ in action~\eqref{eq:monopole_action_Q}. Varying the resulting action with respect to $H_1$ and $H_2$ yields a system:
    \begin{subequations}
    \label{eq:H1_H2_eom}
    \begin{align}
        2 \mathcal{M}_1 H_1 + \mathcal{M}_{12} H_2 &= \frac{1}{2} \left( \frac{n_{13} Q}{n_0} \right)' + \frac{n_2 - n_{13}' + k_1}{2 n_0} Q, \label{eq:eom_H1} \\[6pt]
        \mathcal{M}_{12} H_1 + 2 \mathcal{M}_2 H_2 &= -\frac{1}{2} \partial_t \left( \frac{n_3 Q}{n_0} \right) -\frac{1}{2} \left( \frac{a_3 Q}{n_0} \right)' + \frac{a_7 + k_2}{2 n_0} Q. \label{eq:eom_H2}
    \end{align}
    \end{subequations}
    Solving the linear algebraic system~\eqref{eq:H1_H2_eom} for $H_1$ and $H_2$ and substituting the solution back into Eq.~\eqref{eq:monopole_action_Q}, we arrive at the quadratic action for the single propagating scalar degree of freedom $Q$:
    \begin{equation}
        S^{(2)}_{\ell=0} = \int \mathrm{d}t \, \mathrm{d}r \, \Bigg[ \mathcal{K}^{tt} \dot{Q}^2 + \mathcal{K}^{tr} \dot{Q} Q' - \mathcal{K}^{rr} (Q')^2 - \mathcal{V}_0 Q^2 \Bigg],
        \label{eq:monopole_final_action}
    \end{equation}
    where $\mathcal{V}_0(t,r)$ is an effective potential whose explicit form is irrelevant for the stability analysis of high-energy modes and is therefore omitted. Let us note that in what follows we focus on high-energy modes since these perturbations are able to quickly destroy the background solution, hence, they are the most harmful. 
    The 
    coefficients $\mathcal{K}^{tt}$, $\mathcal{K}^{tr}$, and $\mathcal{K}^{rr}$ entering the action~\eqref{eq:monopole_final_action} can be written in a compact form as follows:
    \begin{equation}
        \mathcal{K}^{tt} = -\frac{\sqrt{A} B^{5/2} \mathcal{T}^4}{2 \mathcal{S}}\Lambda_2, \qquad
        \mathcal{K}^{tr} = -\frac{\sqrt{A} B^{5/2} \mathcal{T}^4 }{\mathcal{S}}\Lambda_3, \qquad
        \mathcal{K}^{rr} = \frac{A^{3/2} B^{7/2} \mathcal{T}^4 }{2 \mathcal{S}}\Lambda_1,
        \label{eq:kinetic_coeffs}
    \end{equation}
    where the common denominator reads
    \begin{equation}
        \mathcal{S} \equiv \left( \Lambda_3^2 - A B \Lambda_1 \Lambda_2 \right) (\dot{\pi}J' - \pi'\dot{J})^2 \left[ \mathcal{T} \dot{B} - B \left( 2 \mathcal{F} \dot{J}^2 - 2 \Gamma_2 J \dot{J} \dot{\pi} + \Sigma_2 J^2 \dot{\pi}^2 \right) \right]^2,
        \label{eq:common_denominator}
    \end{equation}
     and the background functions $\Lambda_i$ and $\mathcal{T}$ are defined as:
    \begin{subequations}
    \label{eq:Lambda_definitions}
    \begin{align}
        \Lambda_1 &= 2 \mathcal{H} J^2 \Gamma_1 \Xi - \mathcal{G} \Xi^2 + 2 \mathcal{H}^2 J^4 \Sigma_1, \label{eq:Lambda1} \\
        \Lambda_2 &= 2 \mathcal{H} J^2 \Gamma_2 \Xi + \mathcal{F} \Xi^2 + 2 \mathcal{H}^2 J^4 \Sigma_2, \label{eq:Lambda2} \\
        \Lambda_3 &= 2 \mathcal{H}^2 J^4 \Sigma_3 - 4 \mathcal{H} J^2 \Gamma_3 \Xi + B \mathcal{D} \Xi^2, \label{eq:Lambda3}\\
        \mathcal{T} &= 2 \mathcal{H} J \dot{J} + \Xi \dot{\pi}.\label{eq:standartT}
    \end{align}
    \end{subequations}
    Here, the coefficients $\mathcal{F}, \mathcal{G}, \mathcal{H}$, and $\mathcal{D}$ are inherited from the odd-parity sector (see Eqs.~\eqref{eq:odd_coeffs}), while the remaining auxiliary coefficients $\Xi, \Gamma_i$, and $\Sigma_i$ are expressed in terms of the Lagrangian functions via the background equations of motion $\mathcal{E}^{(\mu\nu)} = 0$ and $\mathcal{E}^{(\pi)} = 0$ (see Appendix~\ref{app:background_eom} for explicit expressions):
    \begin{subequations}
    \label{eq:auxiliary_coeffs_EoM}
    \begin{align}
        \Sigma_1 &= \frac{1}{B} \frac{\partial \mathcal{E}^{(\pi)}}{\partial \pi''}, \qquad
        \Sigma_2 = A \frac{\partial \mathcal{E}^{(\pi)}}{\partial \ddot{\pi}}, \qquad
        \Sigma_3 = \frac{A}{2} \frac{\partial \mathcal{E}^{(\pi)}}{\partial \dot{\pi}'}, \label{eq:Sigma_def} \\[6pt]
        \Gamma_1 &= -\frac{2}{B} \frac{\partial \mathcal{E}^{(22)}}{\partial \pi''}, \qquad
        \Gamma_2 = -2A \frac{\partial \mathcal{E}^{(22)}}{\partial \ddot{\pi}}, \qquad
        \Gamma_3 = \frac{A}{2} \frac{\partial \mathcal{E}^{(22)}}{\partial \dot{\pi}'}, \label{eq:Gamma_def} \\[6pt]
        \Xi &= -\frac{J^2}{B} \frac{\partial\mathcal{E}^{(00)}}{\partial \pi''}. \label{eq:Xi_def}
    \end{align}
    \end{subequations}
    For the shift-symmetric subclass with $G_5 = 0$, which underlies most known exact black hole and compact object solutions, the explicit expressions for $\Xi$, $\Gamma_i$, and $\Sigma_i$ are provided in Appendix~\ref{app:auxiliary_shift}.
   Notice that in the static limit, one identically has $\mathcal{D} = \Sigma_3 = \Gamma_3 = \mathcal{T} = 0$, which immediately yields $\Lambda_3 = 0$ via Eq.~\eqref{eq:Lambda3}. The rest of the coefficients in Eqs.~\eqref{eq:auxiliary_coeffs_EoM} recover their definitions introduced in the static, spherically symmetric setting in Ref.~\cite{Kobayashi2014_static}.

    \subsection{High-energy stability conditions and radial propagation speeds}
    \label{sec:monopole_stability}
        
    We now examine the linear stability of the monopole perturbations in the high-momentum regime. In this limit, spacetime variations of the perturbation field $Q(t,r)$ occur on characteristic scales much shorter than those of the background metric functions and the scalar field profile $\pi(t,r)$. Consequently, neglecting the effective potential $\mathcal{V}_0$ as well as lower-order derivative terms, and adopting the plane-wave ansatz $Q(t,r) \propto e^{i(\omega t + k r)}$, we obtain the high-energy dispersion relation:
    \begin{equation}
        \mathcal{K}^{tt} \omega^2 + \mathcal{K}^{tr} \omega k - \mathcal{K}^{rr} k^2 = 0.
        \label{eq:dispersion_relation}
    \end{equation}
        To ensure the absence of ghost instabilities, the kinetic coefficient must be strictly positive:
    \begin{equation}
        \mathcal{K}^{tt} > 0.
        \label{eq:no_ghost_general}
    \end{equation}
    Meanwhile, the absence of radial gradient instabilities requires the physical propagation speeds to remain real. The physical propagation speeds along the radial direction, $c_r \equiv \frac{1}{\sqrt{AB}} \frac{\omega}{k}$, are determined by the roots of the dispersion relation~\eqref{eq:dispersion_relation}, giving
    \begin{equation}
        c_r^{(\pm)} = \frac{-\mathcal{K}^{tr} \pm \sqrt{(\mathcal{K}^{tr})^2 + 4 \mathcal{K}^{tt} \mathcal{K}^{rr}}}{2 \sqrt{AB} \mathcal{K}^{tt}},
        \label{eq:cr_general}
    \end{equation}
    where $c_r^{(+)}$ and $c_r^{(-)}$ denote the outward and inward radial propagation speeds, respectively. 
    Hence, the speeds~\eqref{eq:cr_general} are real provided that:
    \begin{equation}
        (\mathcal{K}^{tr})^2 + 4 \mathcal{K}^{tt} \mathcal{K}^{rr} > 0.
        \label{eq:gradient_stability_general}
    \end{equation}
    
    Substituting the explicit expressions for the kinetic coefficients~\eqref{eq:kinetic_coeffs} into Eq.~\eqref{eq:cr_general}, the radial propagation speeds reduce to the explicit form:
    \begin{equation}
        c_r^{(\pm)} = \frac{-\Lambda_3 \pm \sqrt{\Lambda_3^2 - A B \Lambda_1 \Lambda_2}}{\sqrt{AB}\Lambda_2}.
        \label{eq:cr_explicit}
    \end{equation}
    In the GR limit, where the degree of freedom associated with the scalar field is absent, Eq.~\eqref{eq:cr_explicit} correctly yields \(c_r^{(\pm)} = 0\).
    
  We are now in a position to formulate the linear stability criteria for the high-energy monopole modes. Considering the explicit expression for $\mathcal{K}^{tt}$, $\mathcal{K}^{rr}$ and $\mathcal{K}^{tr}$ in Eq.~\eqref{eq:kinetic_coeffs} to the no-ghost condition~\eqref{eq:no_ghost_general} and no-radial gradient instabilities constraint~\eqref{eq:gradient_stability_general},
  the complete set of linear stability criteria for high-energy monopole modes in Horndeski theory reads:
    \begin{align}
        \text{No ghosts:} &\quad \Lambda_2 < 0, \label{eq:final_no_ghost} \\
        \text{No radial gradient instabilities:} &\quad \Lambda_3^2 - A B\cdot\Lambda_1 \Lambda_2 > 0. \label{eq:final_no_gradient}
    \end{align}

    \subsection{Consistency checks and limiting cases}
    \label{sec:limiting_cases}
    
    To further validate our general stability criteria~\eqref{eq:final_no_ghost}-\eqref{eq:final_no_gradient} and the derived radial propagation speed~\eqref{eq:cr_explicit} for the monopole mode, we cross-check our results against the existing results for a static and homogeneous cosmological limiting case in the literature.
    \\

  \textbf{Static limit}
  
     To recover the static limit presented in Ref.~\cite{Kobayashi2014_static}, we map our background variables to the notation of Ref.~\cite{Kobayashi2014_static}.\footnote{In the static limit, the background quantities map as $J(r) = r$, $\pi \to \phi$, $F \to K$, and $K \to G_3$.}
    For a static background, the coefficient $\Lambda_3$ vanishes identically (see Eq.~\eqref{eq:Lambda3}), and taking into account the definitions of $\Lambda_1$ and $\Lambda_2$ given in Eqs.~\eqref{eq:Lambda_definitions}, the radial propagation speed in Eq.~\eqref{eq:cr_explicit} simplifies to
    \begin{equation}
        c_r^2(r) = -\frac{\Lambda_1(r)}{\Lambda_2(r)} = \frac{2 \phi'^2\mathcal{H} J^2 \Gamma_1 \Xi - \mathcal{G} \Xi^2 \phi'^2 + 2 \mathcal{H}^2 J^4 \Sigma_1 \phi'^2}{-\phi'^2(2 \mathcal{H} J^2 \Gamma_2 \Xi + \mathcal{F} \Xi^2 + 2 \mathcal{H}^2 J^4 \Sigma_2)}.
        \label{eq:cr2_static}
    \end{equation}
    The coefficients entering the numerator are related to those of Ref.~\cite{Kobayashi2014_static} as 
    \begin{align*}
        \mathcal{H} = \mathcal{H}_{\text{stat}}, \qquad \Xi = \Xi_{\text{stat}}, \qquad \mathcal{G} = \mathcal{G}_{\text{stat}}, \qquad \Gamma_1 = \Gamma_{\text{stat}}, \qquad \Sigma_1 =\frac{\Sigma_{\text{stat}}}{X}=\frac{-2\Sigma_{\text{stat}}}{B\phi'^2}.
    \end{align*} 
     With these identifications, the numerator of the sound-speed in Eq.~\eqref{eq:cr2_static} directly matches the numerator in the corresponding expression of Ref.~\cite{Kobayashi2014_static}. While the coefficients $\Gamma_2$ and $\Sigma_2$ have no analogues in Ref.~\cite{Kobayashi2014_static}, substituting their explicit definitions via the Lagrangian functions (see Eqs.~\eqref{eq:auxiliary_coeffs_EoM}) and using the static $(00)$- and $(11)$-components of equations of motion (see Eqs.~\eqref{eq:EOM_00},~\eqref{eq:EOM_11}), the denominator in Eq.~\eqref{eq:cr2_static} exactly reproduces that of Ref.~\cite{Kobayashi2014_static}. Thus, the radial propagation speed for the monopole mode in the static background is identically recovered.\\

    \textbf{FLRW limit}
   
    Another consistency check is the cosmological limit, obtained by specializing the background~\eqref{eq:bg_metric} to a spatially flat FLRW spacetime in order to reproduce the scalar perturbation propagation speed given in e.g. Ref.~\cite{Kobayashi2011_flrw}.\footnote{In the FLRW limit, the background variables map as $A = 1$, $B = a^{-2}(t)$, $J = a(t)r$, and $\pi = \phi(t)$, with the Lagrangian functions identified as $F \to K$ and $K \to G_3$.}
    In this case, the coefficient $\Lambda_3$ (Eq.~\eqref{eq:Lambda3}) again vanishes identically, so that the propagation speed of monopole perturbations in Eq.~\eqref{eq:cr_explicit} reduces to $c_s^2(t) = -\Lambda_1(t)/\Lambda_2(t)$.
    In contrast to the static limit where the on-shell reduction is required for the denominator, here the situation is reversed: the denominator $\Lambda_2(t)$ directly reproduces the denominator of the corresponding scalar mode speed, whereas the numerator $\Lambda_1(t)$ requires an on-shell substitution. Specifically, upon writing $\Lambda_1$ in terms of Lagrangian function and eliminating the terms $F_X$ and $K_\pi$ via the background $(00)$- and $(11)$-components of Friedmann equations (see Eqs.~\eqref{eq:EOM_00},~\eqref{eq:EOM_11}), our expression coincides with the propagation speed for the scalar mode derived in Ref.~\cite{Kobayashi2011_flrw}.\\
    
    \textbf{Dynamical, spherically symmetric background in the cubic Horndeski subclass}

  It is important to verify whether the monopole ($\ell = 0$) perturbations considered in this work impose any new stability constraints on a dynamical, spherically symmetric background beyond those derived in Ref.~\cite{MironovSharov2025} for the high-multipole regime ($\ell \gg 1$) in the cubic subclass of Horndeski theory. 
  To do so we map the background metric functions to the conventions of Ref.~\cite{MironovSharov2025} via
    \begin{equation}
        A(t,r) \to a^2(t,r), \qquad B(t,r) \to \frac{1}{b^2(t,r)}, \qquad J(t,r) \to c(t,r),
        \label{eq:metric_rules_MS}
    \end{equation}
    and restrict the Horndeski functions to $G_4 \to 1/(2k)$ and $G_5 \to 0$.\footnote{Since Ref.~\cite{MironovSharov2025} adopts an alternative definition of the kinetic term, $X = \partial_\mu\pi\partial^\mu\pi$, derivatives with respect to $X$ acquire corresponding numerical prefactors: $F_X \to 2 F_X$, $F_{XX} \to 4 F_{XX}$, etc.}
    Under these rules, the coefficients of the quadratic action for the multipoles with $\ell \gg 1$ in Ref.~\cite{MironovSharov2025} are expressed in terms of $\Lambda_i$ functions as follows:
    \begin{equation}
        \mathcal{K}^{00}_{\text{MS}} = -\frac{k^2}{4 c^4} \Lambda_2, \qquad
        \mathcal{K}^{tr}_{\text{MS}} = -\frac{k^2 b}{2 a c^4} \Lambda_3, \qquad
        \mathcal{K}^{rr}_{\text{MS}} = \frac{k^2}{4 c^4} \Lambda_1.
        \label{eq:K_coeffs_MS}
    \end{equation}
    The radial propagation speed derived in Ref.~\cite{MironovSharov2025} then evaluates to:
    \begin{equation}
        c_r = \frac{a}{b} \left( \frac{\mathcal{K}^{tr}_{\text{MS}}}{2\mathcal{K}^{00}_{\text{MS}}} \pm \frac{\sqrt{(\mathcal{K}^{tr}_{\text{MS}})^2 + 4 \mathcal{K}^{00}_{\text{MS}}\mathcal{K}^{rr}_{\text{MS}}}}{2\mathcal{K}^{00}_{\text{MS}}} \right) = -\frac{\Lambda_3}{\Lambda_2} \pm \frac{\sqrt{\Lambda_3^2 - \frac{a^2}{b^2} \Lambda_1 \Lambda_2}}{\Lambda_2},
        \label{eq:cr_MS_eval}
    \end{equation}
    which identically matches our general expression~\eqref{eq:cr_explicit}.
    
     Specifically, the no-ghost requirement $\mathcal{K}^{00}_{\text{MS}} > 0$ yields $\Lambda_2 < 0$, while radial gradient stability requires the speed~\eqref{eq:cr_MS_eval} to be real. Therefore, in the cubic subclass of Horndeski theory, the monopole ($\ell = 0$) sector introduces no additional stability restrictions on the dynamical, spherically symmetric background.

\section{Application to Particular Models}
\label{sec:applications}

\subsection{Non-minimally coupled scalar field}

        Let us consider a massless scalar field non-minimally coupled to the Ricci scalar, corresponding to the following choice of the functions in Lagrangian~\eqref{eq:Lagrangians}:
        \begin{equation}
        F = X, \qquad K = 0, \qquad G_4 = f(\pi), \qquad G_5 = 0.
        \end{equation}
        For this theory, the no-ghost condition~\eqref{eq:final_no_ghost} for the even-parity monopole mode reads
        \begin{equation}
        8\, J^4\, f(\pi)\, \bigl(f(\pi) + 3 f_\pi^2\bigr) > 0.
        \end{equation}
        Remarkably, this condition coincides exactly with the ghost-free requirement obtained in Ref.~\cite{Kobayashi2014_static} for a static background, even without taking the static limit. The radial propagation speed of the scalar mode in Eq.~\eqref{eq:cr_explicit} evaluates to
        \begin{equation}
        c_r^2 = 1,
        \end{equation}
        so that the perturbation propagates at the speed of light.

    \subsection{Hairy black hole in shift-symmetric Horndeski gravity}
    \label{sec:example}
        
        As a direct application of our general stability conditions~\eqref{eq:final_no_ghost} and~\eqref{eq:final_no_gradient} for the monopole mode, we consider a static, spherically symmetric black hole solution with linearly time-dependent scalar hair, $\pi(t,r) = qt + \psi(r)$, in the shift- and reflection-symmetric Horndeski theory presented in Ref.~\cite{TretyakovaTakahashi2017}. While such hairy black holes generically suffer from ghost or gradient instabilities near the horizon in the odd-parity sector, this specific configuration was shown to circumvent this pathology and remain stable against odd-parity perturbations. The theory under consideration is defined as follows:
        \begin{equation}
            F(X) = -2\Lambda + 2\eta X, \qquad K(X) = 0, \qquad G_4(X) = \zeta + \beta X + \frac{\gamma}{2} X^2, \qquad G_5(X) = 0,
            \label{eq:ex_model}
        \end{equation}
        with $\zeta > 0$ and $\gamma \neq 0$. 
    The field equations admit an exact Schwarzschild-dS black hole solution:
        \begin{equation}
            A(r) = -\frac{\gamma q^2}{2\beta} f(r), \qquad B(r) = f(r) = 1 - \frac{\mu}{r} - \frac{\eta}{6\beta} r^2, \qquad J(r) = r,
            \label{eq:ex_metric}
        \end{equation}
        with the radial scalar profile and the cosmological constant given by:
        \begin{equation}
            \psi'^2(r) = -\frac{2\beta}{\gamma} \frac{1 - f(r)}{f^2(r)}, \qquad \Lambda = \frac{\eta}{4\beta\gamma}(2\zeta\gamma - 5\beta^2).
            \label{eq:ex_scalar_Lambda}
        \end{equation}
        On this background, the kinetic term remains constant, $X = -\beta/\gamma$, which yields $G_{4X} = \beta + \gamma X \equiv 0$ and eliminates the dangerous near-horizon divergence in the odd-parity sector. The correct metric signature outside the horizon requires $\beta/\gamma < 0$. Furthermore, the absence of ghost and gradient instabilities in the odd-parity sector demands that (see Sec.~III\,A of Ref.~\cite{TretyakovaTakahashi2017}):
        \begin{equation}
            \zeta > \frac{\beta^2}{2\gamma}.
            \label{eq:ex_odd_condition}
        \end{equation}
        To trace how this constraint affects the even-parity sector, we 
        introduce a
        dimensionless ratio:
        \begin{equation}
            u \equiv \frac{\zeta\gamma}{\beta^2}.
            \label{eq:u_def}
        \end{equation}
        In terms of $u$, the odd-parity stability requirement~\eqref{eq:ex_odd_condition} reads:
        \begin{equation}
            u >\frac{1}{2} \quad \text{for } \gamma > 0, \qquad \text{and} \qquad u <0 \quad \text{for } \gamma < 0.
            \label{eq:odd_u_domains}
        \end{equation}
        
        We now apply our even-parity monopole stability criteria~\eqref{eq:final_no_ghost} and~\eqref{eq:final_no_gradient} to the background given by Eqs.~\eqref{eq:ex_metric} and~\eqref{eq:ex_scalar_Lambda}. Eliminating the mass parameter $\mu$ via the horizon relation $\mu = (1 - f)r - \frac{\eta}{6\beta}r^3$, the conditions ensuring the absence of radial gradient and ghost instabilities evaluate to:
        \begin{align}
            \text{No radial gradient instabilities:} & \quad \frac{\eta}{\gamma} (9 - 2u)(13 + 6u) > 0, \label{eq:ex_grad} \\
            \text{No ghosts:} & \quad (2u - 1) \left[ 4\beta (1 - f(r))^2 (9 - 2u) + (13 + 6u)\eta r^2 \right] < 0. \label{eq:ex_ghost}
        \end{align}
        According to Eq.~\eqref{eq:ex_metric} the requirement $\eta/\beta > 0$ corresponds to asymptotically de Sitter spacetime. Combining the odd-parity conditions~\eqref{eq:odd_u_domains} with the even-parity constraints~\eqref{eq:ex_grad} and~\eqref{eq:ex_ghost} for the monopole mode, one obtains the resulting linear stability criteria, which are summarized in Table~\ref{tab:stability_summary}.
        
        \begin{table}[h!]
        \centering
        \small
        \setlength{\tabcolsep}{8pt}
        \renewcommand{\arraystretch}{1.25}
        \begin{tabular}{lcc}
        \hline\hline
        Couplings & Radial Gradient Stability & Ghost-Free Condition \\
        \hline
        $\gamma > 0, \, \beta < 0, \, \eta < 0$ & $u > \frac{9}{2}$ & $4|\beta|(1-f)^2(2u-9) < (13+6u)|\eta| r^2$ \\
        $\gamma < 0, \, \beta > 0, \, \eta > 0$ & $u < -\frac{13}{6}$ & $4\beta(1-f)^2(9-2u) > |13+6u|\eta r^2$ \\
        \hline\hline
        \end{tabular}
        \caption{Classification of linear stability constraints for even-parity monopole perturbations on the hairy black hole background~\eqref{eq:ex_metric}.}
        \label{tab:stability_summary}
        \end{table}
        
   Thus, the monopole mode further restricts the parameter space originally allowed by the odd-parity analysis. Since perturbations associated with the scalar field were not addressed in Ref.~\cite{TretyakovaTakahashi2017}, the stability criteria derived in this work impose additional constraints on top of those obtained in that paper.

\section{Discussion and Conclusions}
\label{sec:conclusions}

In this paper, we have studied the even-parity monopole ($\ell = 0$) perturbations about an arbitrary time-dependent, spherically symmetric background in full Horndeski theory~\eqref{eq:Lagrangians}. Our goal was to constrain the degree of freedom associated with the scalar field, which does not enter the odd-parity sector and therefore remains unconstrained by the stability conditions of Ref.~\cite{oddSector}. 
 
 We have obtained the quadratic action~\eqref{eq:monopole_final_action} for a single master variable $Q$, which describes the only dynamical degree of freedom in the monopole sector on backgrounds where both the metric functions and the scalar field depend on time and radius. Based on this action, we have formulated the conditions for the absence of ghosts and radial gradient instabilities for high-energy monopole perturbations (see Eqs.~\eqref{eq:final_no_ghost} and~\eqref{eq:final_no_gradient}),
    \begin{align*}
        \text{No ghosts:} & \qquad \Lambda_2 < 0, \\
        \text{No radial gradient instabilities:} & \qquad \Lambda_3^2 - A B \cdot \Lambda_1 \Lambda_2 > 0,
    \end{align*}
    and derived the corresponding radial propagation speeds~\eqref{eq:cr_explicit}. 
    In contrast to the static and cosmological cases, where $\Lambda_3$ vanishes identically, on a generic dynamical background the mixed kinetic term $\mathcal{K}^{tr} \propto \Lambda_3$ is non-zero, so that the outward and inward radial propagation speeds differ in magnitude.
 
    We have restricted our analysis to high-energy modes, since the most dangerous pathologies are ghost and gradient instabilities (see e.g. Ref.~\cite{Rubakov2014} for details). A ghost carries negative kinetic energy, so that its interaction with healthy fields renders the vacuum unstable, while a gradient instability grows at a rate that increases with momentum and therefore destroys the background on arbitrarily short time scales. In contrast to slow tachyonic modes, whose growth rate is set by the background scales, these instabilities are of ultraviolet nature and their absence is a necessary condition for a solution to be physically meaningful. Slow tachyonic instabilities, which are governed by the effective potential $\mathcal{V}_0$ in the action~\eqref{eq:monopole_final_action}, have not been addressed here: whether they are harmful depends on the time scales characteristic of a particular background, since the solution might change significantly during the tachyons' development.
 
    We have verified that, in the appropriate limits, our results reproduce the known propagation speeds for static, spherically symmetric~\cite{Kobayashi2014_static} and spatially flat FLRW~\cite{Kobayashi2011_flrw} backgrounds. In the cubic subclass of Horndeski theory on a dynamical, spherically symmetric background, the monopole conditions coincide with the no-ghost and radial constraints obtained for high-momentum modes in Ref.~\cite{MironovSharov2025}, so that in this subclass the monopole sector imposes no constraints beyond the existing ones. Beyond the cubic subclass, i.e., for a non-constant $G_4$ or a non-zero $G_5$,  our constraints are, to our knowledge, the first stability conditions for the degree of freedom associated with the scalar field on a dynamical, spherically symmetric background~\eqref{eq:bg_metric}.

    We have also demonstrated how the derived monopole stability criteria can be applied within specific subclasses of Horndeski theory, namely,
    a non-minimally coupled scalar field with $G_4 = f(\pi)$ and a shift-symmetric subclass of Horndeski theory admitting a black hole solution with linearly time-dependent scalar hair~\cite{TretyakovaTakahashi2017}, where the constraint from the monopole mode imposes an additional limitation on the model (see Table~\ref{tab:stability_summary}).

    We stress that the conditions obtained in this paper are necessary, but not sufficient, for the stability of the background. By construction, the monopole sector is insensitive to perturbations propagating in the angular directions. Constraining angular propagation requires the analysis of the dipole ($\ell = 1$) and higher multipoles, which on a fully dynamical background~\eqref{eq:bg_metric} is considerably more involved. For $\ell \geq 1$, the angular components of the metric perturbations and of the gauge transformations no longer vanish, and for $\ell \geq 2$ the degree of freedom associated with the scalar field mixes with the even-parity polarization of the graviton, so that integrating out the non-dynamical variables and deriving the master equations become technically demanding. In this respect, the monopole sector is the natural entry point to the even-parity sector on a fully dynamical, spherically symmetric background: it allows all steps of the reduction to be carried out explicitly in full Horndeski theory, and the resulting stability conditions can serve as a benchmark for the analysis of higher multipoles. Finally, since the quadratic action~\eqref{eq:monopole_final_action} has been derived without specifying the background, it may be useful in settings where both inhomogeneity and time dependence are essential, such as gravitational collapse and the dynamical evolution of compact objects.
    
    

\section*{Acknowledgements}
The authors are grateful to Sergey Mironov for numerous and valuable discussions during the preparation of this work.
The work on this project
has been supported by Russian Science Foundation grant № 25-72-00176,
\href{https://rscf.ru/project/25-72-00176/}{  https://rscf.ru/project/25-72-00176/}.

\newpage
\allowdisplaybreaks
\raggedbottom
\appendix
\section{Background Equations of Motion}
\label{app:background_eom}
In this Appendix, we present the explicit analytical form of the background equations of motion. To achieve a compact representation, the terms in each field equation $\mathcal{E}^{(i)} = 0$, with $(i) \in \{(00), (01), (11)\}$, are grouped according to the arbitrary Lagrangian functions $\{F, K, G_4, G_5\}$ and their partial derivatives with respect to the scalar field $\pi$ and the kinetic term $X$:
\begin{equation}
\begin{split}
    \mathcal{E}^{(i)} &= C_{F}^{(i)} F + C_{FX}^{(i)} F_{X}  + C_{K}^{(i)} K + C_{K\pi}^{(i)} K_{\pi} + C_{KX}^{(i)} K_{X} + C_{4}^{(i)} G_4 + C_{4\pi}^{(i)} G_{4\pi} + C_{4X}^{(i)} G_{4X}  + C_{4\pi\pi}^{(i)} G_{4\pi\pi} \\
    &\quad+ C_{4X\pi}^{(i)} G_{4X\pi} + C_{4XX}^{(i)} G_{4XX}  + C_{5\pi}^{(i)} G_{5\pi}  + C_{5X}^{(i)} G_{5X} + C_{5\pi\pi}^{(i)} G_{5\pi\pi} + C_{5X\pi}^{(i)} G_{5X\pi} + C_{5XX}^{(i)} G_{5XX} \,,
\end{split}
\end{equation}
where the coefficients $C_{(\dots)}^{(i)}$ depend exclusively on the metric functions, the scalar field, and their spacetime derivatives.

Furthermore, for each component $(i)$, these coefficients satisfy an algebraic relation:
\begin{equation}
    C_{4X}^{(i)} = - C_{5\pi}^{(i)} - X C_{4}^{(i)} \,.
    \label{eq:app_identity}
\end{equation}
We use this identity to eliminate $C_{4X}^{(i)}$, thereby reducing the redundancy in the equations. The explicit forms of the remaining non-zero coefficients are listed below.

\vspace{0.5cm}

\noindent\textbf{The (00)-component} 

Applying the identities and substituting the compact terms, the $(00)$-component of the equations of motion takes the form:
\begin{equation}\label{eq:EOM_00}
\begin{split}
    \mathcal{E}^{(00)} &= -A F + A(2X + B\pi'^2) F_{X} - 2A(X + B\pi'^2) K_{\pi} + \frac{2B\pi'^2}{J}\mathcal{W} K_{X}  + 2 A B \pi'^2 G_{4\pi\pi} - \frac{8B\pi'^2}{J}\mathcal{W} G_{4X\pi} \\
    &\quad- \frac{2B\pi'^2}{A J^2}\mathcal{W}^2 G_{4XX} + \frac{2B\pi'^2}{J}\mathcal{W} G_{5\pi\pi}  + C_{4}^{(00)} \big(G_4 - X G_{4X}\big)  + C_{4\pi}^{(00)} \big(G_{4\pi} - X K_{X} + 2 X G_{4X\pi}\big) \\
    &\quad + C_{5\pi}^{(00)} \big(G_{5\pi} - G_{4X}\big) + C_{5X\pi}^{(00)} \big(G_{5X\pi} - 2 G_{4XX}\big)  + C_{5X}^{(00)} G_{5X} + C_{5XX}^{(00)} G_{5XX} \,,
\end{split}
\end{equation}
where $\mathcal{W} = \dot{\pi}\dot{J} - A B \pi' J'$, and the unresolved, algebraically extensive coefficients are defined as:
\begin{flalign*}
 C_{4}^{(00)} &=\frac{2}{B J^2} \Big[ A B \left( -1 + B J'^2 + J B' J' + 2 B J J'' \right) + \dot{J} \left( J \dot{B} - B \dot{J} \right) \Big]\,,&\\
 C_{4\pi}^{(00)}&=A \left[ B' \pi' + 2 B \left( \frac{2 J' \pi'}{J} + \pi'' \right) \right] + \left( \frac{\dot{B}}{B} - \frac{4 \dot{J}}{J} \right) \dot{\pi}\,,&\\
 C_{5\pi}^{(00)} &= \frac{2}{B J^2} \bigg[ -A B \Big[ -X \left( -1 + B J'^2 + J B' J' + 2 B     J J'' \right)  + B \pi' \left( (1 + J B' J') \pi' + 2 B J J' \pi'' \right) \Big] \notag & \\
 &\quad + \dot{J} \Big[ J \left( 3 X + B \pi'^2 \right) \dot{B} - B \left( 3 X + 2 B \pi'^2 \right) \dot{J} \Big]  + B \dot{\pi} \Big[ - J J' \pi' \dot{B} + \left( (J B' + 2 B J') \pi' + 2 B J \pi'' \right) \dot{J} \Big] \bigg]\,,&\\
 C_{5X\pi}^{(00)}&=\frac{2}{B J^2} \bigg[ -A B^2 J' \pi' \Big[ B J' \pi' \left( X + B \pi'^2 \right) + J X \left( B' \pi' + 2 B \pi'' \right) \Big]  + \left( 2 X + B \pi'^2 \right) \dot{J} \Big[ J X \dot{B} \notag & \\
&\quad- B \left( X + B \pi'^2 \right) \dot{J} \Big]  + B \dot{\pi} \Big[ 2 B J' \pi' \left( X + B \pi'^2 \right) \dot{J} + J X \left( - J' \pi' \dot{B} + (B' \pi' + 2 B \pi'') \dot{J} \right) \Big] \bigg]\,,&\\
C_{5X}^{(00)} &=\frac{1}{J^2} \bigg\{ A \Big[ -B^2 J'^2 \pi'^2 \left( B' \pi' + 2 B \pi'' \right)  + X \left( \pi' \left( B' (-1 + 3 B J'^2) + 4 B^2 J' J'' \right) + 2 B (-1 + B J'^2) \pi'' \right) \Big] \notag & \\
&\quad + \left( 3 X + B \pi'^2 \right) \dot{J} \left[ 2 J' \pi' \dot{B} - \left( B' \pi' + 2 B \pi'' \right) \dot{J} \right]  + \frac{\dot{\pi}}{A B} \bigg[ - \left( 5 X + B \pi'^2 \right) \dot{B} \dot{J}^2 \notag & \\
&\qquad+ A \Big( \left( X (-1 + B J'^2) - B^2 J'^2 \pi'^2 \right) \dot{B}  + 2 B \left( -X (B' J' + 2 B J'') + B J' \pi' (B' \pi' + 2 B \pi'') \right) \dot{J} \Big) \bigg] \bigg\}\,,&\\
C_{5XX}^{(00)} &=\frac{X}{A B J^2} \bigg\{ A B \Big[ -A B^2 J'^2 \pi'^2 \left( B' \pi' + 2 B \pi'' \right) + \left( 2 X + B \pi'^2 \right) \dot{J} \left( 2 J' \pi' \dot{B} - (B' \pi' + 2 B \pi'') \dot{J} \right) \Big] \notag & \\
&\quad - \dot{\pi} \Big[ \left( 2 X + B \pi'^2 \right) \dot{B} \dot{J}^2 + A B^2 J' \pi' \left( J' \pi' \dot{B} - 2 (B' \pi' + 2 B \pi'') \dot{J} \right) \Big] \bigg\}\,.&
\end{flalign*}

\vspace{0.5cm}

\noindent\textbf{The (01)-component} 

Similarly, for the non-diagonal \((01)\)-component, which is specific to dynamical backgrounds and vanishes in the static limit, we substitute the compact coefficients and apply the exact algebraic identities. By factoring out the common terms and grouping around the heavy coefficients, the equation simplifies to:
\begin{equation}
\begin{split}\label{eq:EOM_01}
    \mathcal{E}^{(01)} &= \pi'\dot{\pi} \left[ F_{X} - 2 K_{\pi} + 2 G_{4\pi\pi} + \frac{2 \mathcal{W}}{A J} \big( K_{X} + G_{5\pi\pi} \big) \right]  + C_{4}^{(01)} \big(G_4 - X G_{4X}\big)  + C_{4\pi}^{(01)} \big(G_{4\pi} - X K_{X}\big) \\
    &\quad + C_{5\pi}^{(01)} \big(G_{5\pi} - G_{4X}\big)  + C_{4X\pi}^{(01)} G_{4X\pi} + C_{4XX}^{(01)} G_{4XX} + C_{5X\pi}^{(01)} G_{5X\pi} + C_{5X}^{(01)} G_{5X} + C_{5XX}^{(01)} G_{5XX} \,,
\end{split}
\end{equation}
where $\mathcal{W} = \dot{\pi}\dot{J} - A B \pi' J'$, and the remaining unresolved, algebraically bulky coefficients are defined as:
\begin{flalign*}
    C_4^{(01)} &= \frac{2}{J} \left( \frac{J' \dot{B}}{B} - \frac{A' \dot{J}}{A} + 2 \dot{J}' \right) \,, &  \\
    C_{4\pi}^{(01)} &= \frac{\pi' \dot{B}}{B} - \frac{A' \dot{\pi}}{A} + 2 \dot{\pi}' \,, &  \\
    C_{5\pi}^{(01)} &= \frac{1}{A^2 B J^2} \bigg\{ - 3 A' B J \dot{J} \dot{\pi}^2 - A^2 B \pi' \Big[ 2 \left( 1 - B J'^2 \right) \dot{\pi} + J \pi' \left( 3 J' \dot{B} + 2 B \dot{J}' \right) + 4 B J J' \dot{\pi}' \Big] \notag & \\
    &\quad + A \Big[ J \dot{B} \dot{\pi} \left( 2 \pi' \dot{J} + J' \dot{\pi} \right) + A' B^2 J \pi' \left( \pi' \dot{J} + 2 J' \dot{\pi} \right) + 2 B \dot{\pi} \left( - \pi' \dot{J}^2 + J (\dot{\pi} \dot{J}' + 2 \dot{J} \dot{\pi}') \right) \Big] \bigg\} \,, & \\
    C_{4X\pi}^{(01)} &= - \pi'^3 \dot{B} + \frac{\pi' \dot{\pi}^2}{A B J} \left( J \dot{B} - 8 B \dot{J} \right) + B \pi'^2 \left[ \left( \frac{A'}{A} + \frac{8 J'}{J} \right) \dot{\pi} - 2 \dot{\pi}' \right] + \frac{\dot{\pi}^2}{A^2} \left( 2 A \dot{\pi}' - A' \dot{\pi} \right) \,, & \\
    C_{4XX}^{(01)} &= \frac{2}{A^3 B J^2} \left( A B J' \pi' - \dot{J} \dot{\pi} \right) \Big\{ - A' B J \dot{\pi}^3 + A \dot{\pi} \Big[ \pi' \left( A' B^2 J \pi' + (J \dot{B} - B \dot{J}) \dot{\pi} \right) + 2 B J \dot{\pi} \dot{\pi}' \Big] \notag & \\
    &\quad + A^2 B \pi'^2 \Big[ B J' \dot{\pi} - J \left( \pi' \dot{B} + 2 B \dot{\pi}' \right) \Big] \Big\} \,, & \\
    C_{5X\pi}^{(01)} &= \frac{1}{A^3 B J^2} \left( A B J' \pi' - \dot{J} \dot{\pi} \right) \Big\{ A' B J \dot{\pi}^3 - A \dot{\pi} \Big[ \pi' \left( A' B^2 J \pi' + (J \dot{B} - 2 B \dot{J}) \dot{\pi} \right) + 2 B J \dot{\pi} \dot{\pi}' \Big] \notag & \\
    &\quad + A^2 B \pi'^2 \Big[ - 2 B J' \dot{\pi} + J \left( \pi' \dot{B} + 2 B \dot{\pi}' \right) \Big] \Big\} \,, &  \\
    C_{5X}^{(01)} &= \frac{1}{2 A^3 B J^2} \bigg\{ 5 A' B \dot{J}^2 \dot{\pi}^3 + A^3 B \pi'^2 \Big[ \pi' \left( (1 - 5 B J'^2) \dot{B} - 4 B^2 J' \dot{J}' \right) + 2 B \left( 1 - 3 B J'^2 \right) \dot{\pi}' \Big] \notag & \\
    &\quad + A^2 \bigg[ \pi' \Big( - \dot{B} \dot{\pi}^2 + A' B^3 J' \pi' (2 \pi' \dot{J} + 3 J' \dot{\pi}) + B \dot{B} (\pi'^2 \dot{J}^2 + 6 J' \pi' \dot{J} \dot{\pi} + 3 J'^2 \dot{\pi}^2) \notag & \\
    &\qquad\qquad\quad + B^2 \dot{\pi} \big( - A' \pi' + 4 (\pi' \dot{J} + J' \dot{\pi}) \dot{J}' \big) \Big) + 2 B \Big( B \pi'^2 \dot{J}^2 + 4 B J' \pi' \dot{J} \dot{\pi} + (B J'^2 - 1) \dot{\pi}^2 \Big) \dot{\pi}' \bigg] \notag & \\
    &\quad - A \dot{\pi} \Big[ \dot{B} \dot{J} \dot{\pi} (3 \pi' \dot{J} + 2 J' \dot{\pi}) + A' B^2 (3 \pi'^2 \dot{J}^2 + 6 J' \pi' \dot{J} \dot{\pi} + J'^2 \dot{\pi}^2) + B \dot{\pi} \left( - \dot{\pi} (A' - 4 \dot{J} \dot{J}') + 6 \dot{J}^2 \dot{\pi}' \right) \Big] \bigg\} \,, &  \\
    C_{5XX}^{(01)} &= \frac{1}{2 A^4 B J^2} \left( A B J' \pi' - \dot{J} \dot{\pi} \right)^2 \left( A B \pi'^2 - \dot{\pi}^2 \right) \Big[ A \left( \pi' \dot{B} + 2 B \dot{\pi}' \right) - A' B \dot{\pi} \Big] \,. & 
\end{flalign*}

\vspace{0.5cm}

\noindent\textbf{The (11)-component} 

Finally, for the spatial $(11)$-component, we substitute the compact coefficients and the corresponding algebraic relations into the general expansion. By factoring out the repetitive structural combinations, we obtain:
\begin{equation}\label{eq:EOM_11}
\begin{split}
    \mathcal{E}^{(11)} &= \frac{1}{B} F + \pi'^2 \big( F_{X} - K_{\pi} \big) + \frac{\dot{\pi}^2}{A B} \left[ - K_{\pi} + 2 G_{4\pi\pi} + \frac{2 \mathcal{W}}{A J} \big( K_{X} + G_{5\pi\pi} \big) \right]  + C_{4}^{(11)} \big(G_4 - X G_{4X}\big) + C_{5XX}^{(11)} G_{5XX} \\
    &\quad+ C_{4\pi}^{(11)} \big(G_{4\pi} - X K_{X}\big)  + C_{5\pi}^{(11)} \big(G_{5\pi} - G_{4X}\big) + C_{4X\pi}^{(11)} G_{4X\pi} + C_{4XX}^{(11)} G_{4XX}  + C_{5X\pi}^{(11)} G_{5X\pi} + C_{5X}^{(11)} G_{5X} \,,
\end{split}
\end{equation}
where $\mathcal{W} = \dot{\pi}\dot{J} - A B \pi' J'$, and the remaining unresolved bulky coefficients are given by:
\begin{flalign*}
C_4^{(11)} &= \frac{2}{A^2 B J^2} \Big[ A^2 \left( 1 - B J'^2 \right) - J \dot{A} \dot{J} + A \left( \dot{J}^2 - A' B J J' + 2 J \ddot{J} \right) \Big] \,, &  \\
C_{4\pi}^{(11)} &= \frac{1}{A^2 B J} \Big[ -4 A^2 B J' \pi' - J \dot{A} \dot{\pi} + A \left( 4 \dot{J} \dot{\pi} - A' B J \pi' + 2 J \ddot{\pi} \right) \Big] \,, &  \\
C_{4X\pi}^{(11)} &= \frac{1}{A^3 B J} \Big\{ 4 A^3 B^2 J' \pi'^3 + A^2 B \pi' \left[ A' B J \pi'^2 + 4 J' \dot{\pi}^2 - 2 \pi' \left( 2 \dot{J} \dot{\pi} + J \ddot{\pi} \right) \right] \notag & \\
&\quad + A \dot{\pi} \left[ B J \pi' \left( \dot{A} \pi' - A' \dot{\pi} \right) + 2 \dot{\pi} \left( J \ddot{\pi} - 2 \dot{J} \dot{\pi} \right) \right] - J \dot{A} \dot{\pi}^3 \Big\} \,, & \\
C_{4XX}^{(11)} &= \frac{2}{A^4 B J^2} \left( A B J' \pi' - \dot{J} \dot{\pi} \right) \Big\{ A^3 B^2 J' \pi'^3 + A^2 B \pi'^2 \left[ - \dot{J} \dot{\pi} + J \left( A' B \pi' - 2 \ddot{\pi} \right) \right]  + A J \dot{\pi} \left[ B \pi' \left( \dot{A} \pi' - A' \dot{\pi} \right) + 2 \dot{\pi} \ddot{\pi} \right] - J \dot{A} \dot{\pi}^3 \Big\} \,, &  \\
C_{5\pi}^{(11)} &= \frac{1}{A^3 B J^2} \bigg\{ A^3 B \left( 3 B J'^2 - 1 \right) \pi'^2 + A^2 \Big[ 3 A' B^2 J J' \pi'^2 - \dot{\pi}^2 + B J'^2 \dot{\pi}^2  - B \pi'^2 \left( \dot{J}^2 + 2 J \ddot{J} \right) - 4 B J' \pi' \left( \dot{J} \dot{\pi} + J \ddot{\pi} \right) \Big] \notag & \\
&\quad + A \Big[ B J \left( \dot{A} \dot{J} \pi'^2 + 2 (J' \dot{A} - A' \dot{J}) \pi' \dot{\pi} - A' J' \dot{\pi}^2 \right) + \dot{\pi}^2 \left( \dot{J}^2 + 2 J \ddot{J} \right) + 4 J \dot{J} \dot{\pi} \ddot{\pi} \Big]  - 3 J \dot{A} \dot{J} \dot{\pi}^2 \bigg\} \,, &  \\
C_{5X\pi}^{(11)} &= - \frac{1}{A^4 B J^2} \left( A B J' \pi' - \dot{J} \dot{\pi} \right) \Big\{ A^3 B^2 J' \pi'^3 + A^2 B \pi' \left[ A' B J \pi'^2 + J' \dot{\pi}^2 - \pi' \left( \dot{J} \dot{\pi} + 2 J \ddot{\pi} \right) \right] \notag & \\
&\quad + A \dot{\pi} \left[ B J \pi' \left( \dot{A} \pi' - A' \dot{\pi} \right) + \dot{\pi} \left( 2 J \ddot{\pi} - \dot{J} \dot{\pi} \right) \right] - J \dot{A} \dot{\pi}^3 \Big\} \,, &  \\
C_{5XX}^{(11)} &= - \frac{1}{2 A^5 B J^2} \left( A B J' \pi' - \dot{J} \dot{\pi} \right)^2 \left( A B \pi'^2 - \dot{\pi}^2 \right) \Big[ \dot{A} \dot{\pi} + A \left( A' B \pi' - 2 \ddot{\pi} \right) \Big] \,. & \\
C_{5X}^{(11)} &= \frac{1}{2 A^4 B J^2} \bigg\{ 5 \dot{A} \dot{J}^2 \dot{\pi}^3 + A^3 B \pi'^2 \Big[ B \pi' \left( A' (5 B J'^2 - 1) - 4 J' \ddot{J} \right) + 2 \left( 1 - 3 B J'^2 \right) \ddot{\pi} \Big] \notag & \\
&\quad + A^2 \bigg[ B \pi' \Big( - A' B \pi'^2 \dot{J}^2 + 2 B J' \pi' \dot{J} (\dot{A} \pi' - 3 A' \dot{\pi}) + 3 B J'^2 \dot{\pi} (\dot{A} \pi' - A' \dot{\pi}) \notag & \\
&\qquad\qquad\quad + \dot{\pi} \big( \dot{\pi} (A' + 4 J' \ddot{J}) - \pi' (\dot{A} - 4 \dot{J} \ddot{J}) \big) \Big) \notag  + 2 \Big( B \pi'^2 \dot{J}^2 + 4 B J' \pi' \dot{J} \dot{\pi} + (B J'^2 - 1) \dot{\pi}^2 \Big) \ddot{\pi} \bigg] \notag & \\
&\quad - A \dot{\pi} \bigg[ B \Big( 3 \dot{A} \dot{J}^2 \pi'^2 - 3 \dot{J} (A' \dot{J} - 2 J' \dot{A}) \pi' \dot{\pi} + J' (J' \dot{A} - 2 A' \dot{J}) \dot{\pi}^2 \Big) \notag + \dot{\pi} \left( - \dot{\pi} (\dot{A} - 4 \dot{J} \ddot{J}) + 6 \dot{J}^2 \ddot{\pi} \right) \bigg] \bigg\} \,. & 
\end{flalign*}

\vspace{0.5cm}

\noindent\textbf{Bianchi identities and the remaining equations} 

Due to the general covariance of the action, the generalized Bianchi identities ensure that the remaining transverse component $\mathcal{E}^{(22)}$ and the scalar field equation $\mathcal{E}^{(\pi)}$ are not dynamically independent. Instead, they can be explicitly expressed as exact linear combinations of the primary components $\mathcal{E}^{(00)}$, $\mathcal{E}^{(01)}$, and $\mathcal{E}^{(11)}$ and their derivatives:
\begin{flalign}
\mathcal{E}^{(22)} &= \frac{1}{2 J \left( J' \dot{\pi} - \dot{J} \pi' \right)} \Bigg[ 
    \frac{J^2}{2} \left( \frac{A'}{A} \dot{\pi} - \frac{\dot{B}}{B} \pi' \right) \left( \mathcal{E}^{(11)} - \mathcal{E}^{(00)} \right)  + \dot{\pi} \left( J^2 \mathcal{E}^{(11)} \right)' - \pi' \partial_t \left( J^2 \mathcal{E}^{(00)} \right)  - \dot{\pi} \partial_t \left( \frac{2 J^2}{A} \mathcal{E}^{(01)} \right) \notag & \\
&\quad- \pi' \left( 2 B J^2 \mathcal{E}^{(01)} \right)' \notag  + J^2 \left\{ \frac{\dot{\pi}}{B} \partial_t \left( \frac{B}{A} \right) + A \pi' \left( \frac{B}{A} \right)' \right\} \mathcal{E}^{(01)} \Bigg] \,,  \\
\mathcal{E}^{(\pi)} &= \frac{1}{2 \left( J' \dot{\pi} - \dot{J} \pi' \right)} \Bigg\{ 
    2 B \left( J' \mathcal{E}^{(01)\prime} - \dot{J} \mathcal{E}^{(11)\prime} \right) 
    + \frac{2}{A} \left( \dot{J} \dot{\mathcal{E}}^{(01)} - J' \dot{\mathcal{E}}^{(00)} \right) + \mathcal{E}^{(00)} \left[ \frac{J'}{A} \left( \frac{2 \dot{A}}{A} + \frac{\dot{B}}{B} - \frac{4 \dot{J}}{J} \right) - \frac{A' \dot{J}}{A^2} \right] \notag & \\
&\quad + \mathcal{E}^{(01)} \left[ B J' \left( \frac{A'}{A} + \frac{B'}{B} + \frac{4 J'}{J} \right) - \frac{\dot{J}}{A} \left( \frac{\dot{A}}{A} + \frac{\dot{B}}{B} - \frac{4 \dot{J}}{J} \right) \right]  + \mathcal{E}^{(11)} \left[ J' \dot{B} - B \dot{J} \left( \frac{A'}{A} + \frac{2 B'}{B} + \frac{4 J'}{J} \right) \right] \Bigg\} \,. 
\end{flalign}
Therefore, any solution to the primary $(00)$-, $(01)$-, and $(11)$-components automatically satisfies the full dynamics of the system.

\vspace{0.5cm}

    \noindent\textbf{Background equations in effective variables} 
    
    To establish a direct connection with the main text, the primary background equations of motion can be rewritten in a compact form using the effective variables \(\Omega_1\), \(\Omega_2\), \(\mathcal{H}\), and \(\Xi\) introduced in Eqs.~\eqref{eq:Omega_definitions}, \eqref{eq:odd_H}, and \eqref{eq:Xi_def}. In this notation, the explicit form of the \((00)\)-, \((11)\)-, and \((01)\)-components reads:
    \begin{subequations}
    \label{eq:app_background_eqs}
    \begin{align}
        \mathcal{E}^{(00)} &=- \frac{B}{J^2} \Bigg[ \Omega_1 (\pi')^2 + \frac{1}{B} \left( -\Omega_2 J^2 + \mathcal{H} J B' J' + \frac{1}{2} \Xi B' \pi' \right) + 2 \mathcal{H} J J'' + \Xi \pi'' + \frac{\dot{B}}{2 A B^2}\left( 2 \mathcal{H} J \dot{J} + \Xi \dot{\pi} \right) \Bigg], \label{eq:app_E00} \\[8pt]
        \mathcal{E}^{(11)} &= \frac{1}{A J^2} \Bigg[ \Omega_2 A J^2 - \frac{1}{2} B A' \left( 2 \mathcal{H} J J' + \Xi \pi' \right) + \Omega_1 \dot{\pi}^2 - \frac{\dot{A}}{2 A}\left( 2 \mathcal{H} J \dot{J} + \Xi \dot{\pi} \right) + 2 \mathcal{H} J \ddot{J} + \Xi \ddot{\pi} \Bigg], \label{eq:app_E11} \\[8pt]
        \mathcal{E}^{(01)} &= \frac{1}{2 J^2} \Bigg[ 2 \mathcal{H} J \dot{J}' + \frac{\dot{B}}{2 B} \left( 2 \mathcal{H} J J' + \Xi \pi' \right) + \Omega_1 \pi' \dot{\pi} - \frac{A'}{2 A}\left( 2 \mathcal{H} J \dot{J} + \Xi \dot{\pi} \right) + \Xi \dot{\pi}' \Bigg]. \label{eq:app_E01}
    \end{align}
    \end{subequations}
    
    The effective variables $\Omega_1$ and $\Omega_2$ absorb the extensive dependence on the kinetic term $X$ and the higher-order derivatives of the Horndeski Lagrangian functions. Their explicit expressions are given by:
    \begin{subequations}
    \label{eq:Omega_definitions}
    \begin{flalign}
        \Omega_1 &= J^2 \big( F_X - 2 K_\pi + 2 G_{4\pi\pi} \big) + 2 \left( 1 - B J'^2 + \frac{\dot{J}^2}{A} \right) \big( G_{4X} - G_{5\pi} \big) \nonumber & \\
        &\quad + \frac{2 J (\dot{\pi}\dot{J} - A B \pi' J')}{A} \big( K_X - 4 G_{4X\pi} + G_{5\pi\pi} \big) + \frac{2 (\dot{\pi}\dot{J} - A B \pi' J')^2}{A^2} \big( G_{4XX} - G_{5X\pi} \big) \,, & \label{eq:Omega1} \\[10pt]
        \Omega_2 &= F - 2 X \big( F_X - K_\pi \big) + \frac{2 G_4}{J^2} \left( 1 - B J'^2 + \frac{\dot{J}^2}{A} \right) + \frac{4 (\dot{\pi}\dot{J} - A B \pi' J')}{A J} G_{4\pi}  - \frac{2 X}{J^2} \left( 1 - B J'^2 + \frac{\dot{J}^2}{A} \right) \big( 2 G_{4X} - G_{5\pi} \big) \nonumber & \\
        &\quad- \frac{2 (\dot{\pi}\dot{J} - A B \pi' J')^2}{A^2 J^2} \big( G_{4X} - G_{5\pi} \big) - \frac{4 X (\dot{\pi}\dot{J} - A B \pi' J')}{A J} \big( K_X - 2 G_{4X\pi} \big) - \frac{2 X (\dot{\pi}\dot{J} - A B \pi' J')^2}{A^2 J^2} \big( 2 G_{4XX} - G_{5X\pi} \big) \,. & \label{eq:Omega2}
    \end{flalign}
    \end{subequations}
    Here, $X$ is the background canonical kinetic term defined in Eq.~\eqref{eq:kinetic_term}.

\section{Coefficients of the Monopole Quadratic Action}
\label{app:monopole_coeffs}

The explicit expressions for the background-dependent coefficients entering the quadratic monopole action~\eqref{eq:monopole_action} read:
\begin{flalign*}
n_0 &= -\frac{1}{4 \sqrt{A} \sqrt{B}} \left[ -\frac{\mathcal{T} \dot{B}}{B} + 2 \mathcal{F} \dot{J}^2 - 2 \Gamma_2 J \dot{J} \dot{\pi} + \Sigma_2 J^2 \dot{\pi}^2 \right], &\\
a_7 &= -\left( \frac{1}{2} \sqrt{A} \sqrt{B} \mathcal{R} \right)' - \frac{\mathrm{d}}{\mathrm{d}t} \left( \frac{\mathcal{T}}{2 \sqrt{A} \sqrt{B}} \right) 
- \frac{1}{2 \sqrt{A} \sqrt{B}} \left[ 2 A J^2 \Omega_2 - B \mathcal{R} A' - 2 \mathcal{G} \dot{J}^2 - 2 \Gamma_1 J \dot{J} \dot{\pi} + \Sigma_1 J^2 \dot{\pi}^2 \right], &\\
b_1 &= -\frac{1}{A} \frac{\mathrm{d}}{\mathrm{d}t} \left( \frac{\sqrt{B} \mathcal{T}}{\sqrt{A}} \right) 
- \frac{\sqrt{B}}{A^{3/2}} \left[ -2 \mathcal{G} \dot{J}^2 + J \dot{\pi} \left( -2 \Gamma_1 \dot{J} + J \Sigma_1 \dot{\pi} \right) \right], &\\
c_6 &= \frac{\sqrt{B}}{4 \sqrt{A}} \left[ \mathcal{R} A' + A \left( 2 \mathcal{G} (J')^2 + J \pi' \left( 2 \Gamma_1 J' - J \Sigma_1 \pi' \right) \right) \right], &\\
a_3 &=- \frac{1}{2} \sqrt{A} \sqrt{B} \mathcal{R}, &\\
n_{13} &=- \frac{\sqrt{B} \mathcal{T}}{\sqrt{A}}, &\\
n_3 &=- \frac{\mathcal{T}}{2 \sqrt{A} \sqrt{B}}, &\\
b_4 &= \frac{\sqrt{B} \mathcal{R}}{\sqrt{A}} ,&\\
n_9 &= -\frac{1}{2 A^{3/2}} \left[ \sqrt{B} \mathcal{T} A' - 2 A \left( \frac{\mathrm{d}}{\mathrm{d}t}(\sqrt{B} \mathcal{R}) - \sqrt{B} \left( 2 \mathcal{G} J' \dot{J} + J \left[ -J \Sigma_1 \pi' \dot{\pi} + \Gamma_1 (\pi' \dot{J} + J' \dot{\pi}) \right] \right) \right) \right] ,&\\
n_2 &= -\frac{1}{2 \sqrt{A} \sqrt{B}} \left[ 2 \sqrt{A} B \left( \frac{\mathcal{T}}{\sqrt{A}} \right)' + \mathcal{R} \dot{B} + 2 B \left( -2 \mathcal{F} J' \dot{J} + J \left[ -J \Sigma_2 \pi' \dot{\pi} + \Gamma_2 (\pi' \dot{J} + J' \dot{\pi}) \right] \right) \right], &
\end{flalign*}
where $\mathcal{R} = 2 \mathcal{H} J J' + \Xi \pi'$.

\section{Explicit Auxiliary Coefficients in Shift-Symmetric Horndeski Gravity}
\label{app:auxiliary_shift}

While the general expressions for the auxiliary coefficients $\Sigma_i$, $\Gamma_i$, and $\Xi$ in full Horndeski theory are rather bulky, they simplify considerably in shift-symmetric theories where the Lagrangian functions depend on the scalar field only through its derivatives. In this shift-symmetric subclass with $G_5 = 0$ ($F=F(X)$, $K=K(X)$, and $G_4=G_4(X)$), which is widely employed to construct exact hairy black hole and compact object configurations, the explicit form of all background auxiliary functions $\Xi$, $\Gamma_i$, and $\Sigma_i$ reads as follows:
\begin{subequations}
\label{eq:auxiliary_functions_shift}
\begin{flalign}
    \Xi &= J^2 \left[ -2 X K_X - \frac{4(\dot{\pi}\dot{J} - A B \pi' J')}{A J} \Big( G_{4X} + 2 X G_{4XX} \Big) \right] \,, &  \\
    \Gamma_1 &= 2 B (\pi')^2 K_X - \frac{4(\dot{\pi}\dot{J} - A B \pi' J')}{A J} \Big( G_{4X} - B (\pi')^2 G_{4XX} \Big) - 2 \Big( G_{4X} + 2 X G_{4XX} \Big) \left( \frac{2\ddot{\pi}}{A} - \frac{\dot{A}\dot{\pi}}{A^2} - \frac{B A'\pi'}{A} \right) \,, &  \\
    \Gamma_2 &= \frac{2 \dot{\pi}^2}{A} K_X + \frac{4(\dot{\pi}\dot{J} - A B \pi' J')}{A J} \left( G_{4X} + \frac{\dot{\pi}^2}{A} G_{4XX} \right) \nonumber  - 2 \Big( G_{4X} + 2 X G_{4XX} \Big) \left( 2 B \pi'' + B'\pi' + \frac{\dot{B}\dot{\pi}}{A B} \right) \,, &  \\
    \Gamma_3 &= B \dot{\pi} \pi' \left( K_X + \frac{2(\dot{\pi}\dot{J} - A B \pi' J')}{A J} G_{4XX} \right)  - B \Big( G_{4X} + 2 X G_{4XX} \Big) \left( 2 \dot{\pi}' - \frac{A'\dot{\pi}}{A} + \frac{\dot{B}\pi'}{B} \right) \,, &  \\
    \Sigma_1 &= F_X - B (\pi')^2 F_{XX}  + \left( \frac{2\ddot{\pi}}{A} - \frac{\dot{A}\dot{\pi}}{A^2} - \frac{B A'\pi'}{A} \right) \left[ K_X + X K_{XX} + \frac{2(\dot{\pi}\dot{J} - A B \pi' J')}{A J} \Big( 3 G_{4XX} + 2 X G_{4XXX} \Big) \right] \nonumber & \\
    &\quad + \frac{2(\dot{\pi}\dot{J} - A B \pi' J')}{A J} \Big[ 2 K_X - B (\pi')^2 K_{XX} \Big] + \frac{2}{J} \Big( G_{4X} + 2 X G_{4XX} \Big) \left( \frac{2\ddot{J}}{A} - \frac{\dot{A}\dot{J}}{A^2} - \frac{B A' J'}{A} \right) \nonumber & \\
    &\quad + \frac{2}{J^2}\left(1 - B (J')^2 + \frac{\dot{J}^2}{A}\right) \Big[ G_{4X} - B (\pi')^2 G_{4XX} \Big]  + \frac{2(\dot{\pi}\dot{J} - A B \pi' J')^2}{A^2 J^2} \Big[ 3 G_{4XX} - B (\pi')^2 G_{4XXX} \Big] \,, &  \\
    \Sigma_2 &= - \left( F_X + \frac{\dot{\pi}^2}{A} F_{XX} \right)  + \left( 2 B \pi'' + B'\pi' + \frac{\dot{B}\dot{\pi}}{A B} \right) \left[ K_X + X K_{XX} + \frac{2(\dot{\pi}\dot{J} - A B \pi' J')}{A J} \Big( 3 G_{4XX} + 2 X G_{4XXX} \Big) \right] \nonumber & \\
    &\quad - \frac{2(\dot{\pi}\dot{J} - A B \pi' J')}{A J} \left[ 2 K_X + \frac{\dot{\pi}^2}{A} K_{XX} \right] + \frac{2}{J} \Big( G_{4X} + 2 X G_{4XX} \Big) \left( 2 B J'' + B' J' + \frac{\dot{B}\dot{J}}{A B} \right) \nonumber & \\
    &\quad - \frac{2}{J^2}\left(1 - B (J')^2 + \frac{\dot{J}^2}{A}\right) \left[ G_{4X} + \frac{\dot{\pi}^2}{A} G_{4XX} \right] - \frac{2(\dot{\pi}\dot{J} - A B \pi' J')^2}{A^2 J^2} \left[ 3 G_{4XX} + \frac{\dot{\pi}^2}{A} G_{4XXX} \right] \,, &  \\
    \Sigma_3 &= B \dot{\pi} \pi' \Bigg[ F_{XX} + \frac{2(\dot{\pi}\dot{J} - A B \pi' J')}{A J} K_{XX} + \frac{2}{J^2}\left(1 - B (J')^2 + \frac{\dot{J}^2}{A}\right) G_{4XX} + \frac{2(\dot{\pi}\dot{J} - A B \pi' J')^2}{A^2 J^2} G_{4XXX} \Bigg] \nonumber & \\
    &\quad - B \left( 2 \dot{\pi}' - \frac{A'\dot{\pi}}{A} + \frac{\dot{B}\pi'}{B} \right) \left[ K_X + X K_{XX} + \frac{2(\dot{\pi}\dot{J} - A B \pi' J')}{A J} \Big( 3 G_{4XX} + 2 X G_{4XXX} \Big) \right] \nonumber & \\
    &\quad - \frac{2 B}{J} \Big( G_{4X} + 2 X G_{4XX} \Big) \left( 2 \dot{J}' - \frac{A'\dot{J}}{A} + \frac{\dot{B}J'}{B} \right) \,. & 
\end{flalign}
\end{subequations}



\bibliographystyle{JHEP}
\bibliography{flavNEW}

\providecommand{\href}[2]{#2}\begingroup\raggedright\begin{thebibliography}{10}

\bibitem{Horndeski1974}
G.~W. Horndeski, \emph{{Second-order scalar-tensor field equations in a four-dimensional space}}, \href{https://doi.org/10.1007/BF01807638}{\emph{Int. J. Theor. Phys.} {\bfseries 10} (1974) 363}.

\bibitem{Nicolis2009}
A.~Nicolis, R.~Rattazzi and E.~Trincherini, \emph{{The Galileon as a local modification of gravity}}, \href{https://doi.org/10.1103/PhysRevD.79.064036}{\emph{Phys. Rev. D} {\bfseries 79} (2009) 064036} [\href{https://arxiv.org/abs/0811.2197}{{\ttfamily 0811.2197}}].

\bibitem{Deffayet2009generalized}
C.~Deffayet, S.~Deser and G.~Esposito-Far{\`e}se, \emph{{Generalized Galileons: All scalar models whose curved background extensions maintain second-order field equations and stress-tensors}}, \href{https://doi.org/10.1103/PhysRevD.80.064015}{\emph{Phys. Rev. D} {\bfseries 80} (2009) 064015} [\href{https://arxiv.org/abs/0906.1967}{{\ttfamily 0906.1967}}].

\bibitem{Deffayet2011}
C.~Deffayet, X.~Gao, D.~A. Steer and G.~Zahariade, \emph{{From k-essence to generalised Galileons}}, \href{https://doi.org/10.1103/PhysRevD.84.064039}{\emph{Phys. Rev. D} {\bfseries 84} (2011) 024012} [\href{https://arxiv.org/abs/1103.3260}{{\ttfamily 1103.3260}}].

\bibitem{Kobayashi2011_flrw}
T.~Kobayashi, M.~Yamaguchi and J.~Yokoyama, \emph{{Generalized G-inflation: Inflation with the most general second-order field equations}}, \href{https://doi.org/10.1143/PTP.126.511}{\emph{Prog. Theor. Phys.} {\bfseries 126} (2011) 511} [\href{https://arxiv.org/abs/1105.5723}{{\ttfamily 1105.5723}}].

\bibitem{Zumalacarregui2014}
M.~Zumalac{\'a}rregui and J.~Garc{\'i}a-Bellido, \emph{{Transforming gravity: from derivative couplings to matter to second-order scalar-tensor theories beyond the Horndeski Lagrangian}}, \href{https://doi.org/10.1103/PhysRevD.89.064046}{\emph{Phys. Rev. D} {\bfseries 89} (2014) 064046} [\href{https://arxiv.org/abs/1308.4685}{{\ttfamily 1308.4685}}].

\bibitem{Gleyzes2015}
J.~Gleyzes, D.~Langlois, F.~Piazza and F.~Vernizzi, \emph{{Healthy theories beyond Horndeski}}, \href{https://doi.org/10.1103/PhysRevLett.114.211101}{\emph{Phys. Rev. Lett.} {\bfseries 114} (2015) 211101} [\href{https://arxiv.org/abs/1404.6495}{{\ttfamily 1404.6495}}].

\bibitem{Gleyzes2015exploring}
J.~Gleyzes, D.~Langlois, F.~Piazza and F.~Vernizzi, \emph{{Exploring gravitational theories beyond Horndeski}}, \href{https://doi.org/10.1088/1475-7516/2015/02/018}{\emph{JCAP} {\bfseries 02} (2015) 018} [\href{https://arxiv.org/abs/1408.1952}{{\ttfamily 1408.1952}}].

\bibitem{LangloisNoui2016}
D.~Langlois and K.~Noui, \emph{{Degenerate higher derivative theories beyond Horndeski: evading the Ostrogradski instability}}, \href{https://doi.org/10.1088/1475-7516/2016/02/034}{\emph{JCAP} {\bfseries 02} (2016) 034} [\href{https://arxiv.org/abs/1510.06930}{{\ttfamily 1510.06930}}].

\bibitem{Crisostomi2016}
M.~Crisostomi, K.~Koyama and G.~Tasinato, \emph{{Extended Scalar-Tensor Theories of Gravity}}, \href{https://doi.org/10.1088/1475-7516/2016/04/044}{\emph{JCAP} {\bfseries 04} (2016) 044} [\href{https://arxiv.org/abs/1602.03119}{{\ttfamily 1602.03119}}].

\bibitem{BenAchour2016disformal}
J.~Ben~Achour, D.~Langlois and K.~Noui, \emph{{Degenerate higher order scalar-tensor theories beyond Horndeski and disformal transformations}}, \href{https://doi.org/10.1103/PhysRevD.93.124005}{\emph{Phys. Rev. D} {\bfseries 93} (2016) 124005} [\href{https://arxiv.org/abs/1602.08398}{{\ttfamily 1602.08398}}].

\bibitem{BenAchour2016cubic}
J.~Ben~Achour, M.~Crisostomi, K.~Koyama, D.~Langlois, K.~Noui and G.~Tasinato, \emph{{Degenerate higher order scalar-tensor theories beyond Horndeski up to cubic order}}, \href{https://doi.org/10.1007/JHEP12(2016)100}{\emph{JHEP} {\bfseries 12} (2016) 100} [\href{https://arxiv.org/abs/1608.08135}{{\ttfamily 1608.08135}}].

\bibitem{Kobayashi2019review}
T.~Kobayashi, \emph{{Horndeski theory and beyond: a review}}, \href{https://doi.org/10.1088/1361-6633/ab2429}{\emph{Rept. Prog. Phys.} {\bfseries 82} (2019) 086901} [\href{https://arxiv.org/abs/1901.07183}{{\ttfamily 1901.07183}}].

\bibitem{Langlois2019review}
D.~Langlois, \emph{{Dark energy and modified gravity in degenerate higher-order scalar-tensor (DHOST) theories: A review}}, \href{https://doi.org/10.1142/S0218271819420069}{\emph{Int. J. Mod. Phys. D} {\bfseries 28} (2019) 1942006} [\href{https://arxiv.org/abs/1811.06271}{{\ttfamily 1811.06271}}].

\bibitem{Rubakov2014}
V.~A. Rubakov, \emph{{The Null Energy Condition and its violation}}, \href{https://doi.org/10.3367/UFNe.0184.201402b.0137}{\emph{Phys. Usp.} {\bfseries 57} (2014) 128} [\href{https://arxiv.org/abs/1401.4024}{{\ttfamily 1401.4024}}].

\bibitem{Creminelli2010genesis}
P.~Creminelli, A.~Nicolis and E.~Trincherini, \emph{{Galilean Genesis: An Alternative to inflation}}, \href{https://doi.org/10.1088/1475-7516/2010/11/021}{\emph{JCAP} {\bfseries 11} (2010) 021} [\href{https://arxiv.org/abs/1007.0027}{{\ttfamily 1007.0027}}].

\bibitem{Easson2011}
D.~A. Easson, I.~Sawicki and A.~Vikman, \emph{{G-Bounce}}, \href{https://doi.org/10.1088/1475-7516/2011/11/021}{\emph{JCAP} {\bfseries 11} (2011) 021} [\href{https://arxiv.org/abs/1109.1047}{{\ttfamily 1109.1047}}].

\bibitem{Heisenberg:2018vsk}
L.~Heisenberg, \emph{{A systematic approach to generalisations of General Relativity and their cosmological implications}}, \href{https://doi.org/10.1016/j.physrep.2018.11.006}{\emph{Phys. Rept.} {\bfseries 796} (2019) 1} [\href{https://arxiv.org/abs/1807.01725}{{\ttfamily 1807.01725}}].

\bibitem{MironovVolkova2024review}
V.~E. Volkova and S.~A. Mironov, \emph{{Nonsingular cosmological scenarios in scalar-tensor theories and their stability}}, \href{https://doi.org/10.3367/UFNr.2024.12.039826}{\emph{Usp. Fiz. Nauk} {\bfseries 195} (2025) 163} [\href{https://arxiv.org/abs/2409.16108}{{\ttfamily 2409.16108}}].

\bibitem{Deffayet2010KGB}
C.~Deffayet, O.~Pujolas, I.~Sawicki and A.~Vikman, \emph{{Imperfect Dark Energy from Kinetic Gravity Braiding}}, \href{https://doi.org/10.1088/1475-7516/2010/10/026}{\emph{JCAP} {\bfseries 10} (2010) 026} [\href{https://arxiv.org/abs/1008.0048}{{\ttfamily 1008.0048}}].

\bibitem{DeFeliceTsujikawa2012}
A.~De~Felice and S.~Tsujikawa, \emph{{Conditions for the cosmological viability of the most general scalar-tensor theories and their applications to extended Galileon dark energy models}}, \href{https://doi.org/10.1088/1475-7516/2012/02/007}{\emph{JCAP} {\bfseries 02} (2012) 007} [\href{https://arxiv.org/abs/1110.3878}{{\ttfamily 1110.3878}}].

\bibitem{Gleyzes2013}
J.~Gleyzes, D.~Langlois, F.~Piazza and F.~Vernizzi, \emph{{Essential building blocks of dark energy}}, \href{https://doi.org/10.1088/1475-7516/2013/08/025}{\emph{JCAP} {\bfseries 08} (2013) 025} [\href{https://arxiv.org/abs/1304.4840}{{\ttfamily 1304.4840}}].

\bibitem{KaseTsujikawa2019}
R.~Kase and S.~Tsujikawa, \emph{{Dark energy in Horndeski theories after GW170817: A review}}, \href{https://doi.org/10.1142/S021827181942005X}{\emph{Int. J. Mod. Phys. D} {\bfseries 28} (2019) 1942005} [\href{https://arxiv.org/abs/1809.08735}{{\ttfamily 1809.08735}}].

\bibitem{LIGO2017GW170817}
{\scshape LIGO Scientific, Virgo, Fermi-GBM, INTEGRAL} collaboration, \emph{{Gravitational Waves and Gamma-rays from a Binary Neutron Star Merger: GW170817 and GRB 170817A}}, \href{https://doi.org/10.3847/2041-8213/aa920c}{\emph{Astrophys. J. Lett.} {\bfseries 848} (2017) L13} [\href{https://arxiv.org/abs/1710.05834}{{\ttfamily 1710.05834}}].

\bibitem{Ezquiaga2017}
J.~M. Ezquiaga and M.~Zumalac{\'a}rregui, \emph{{Dark Energy After GW170817: Dead Ends and the Road Ahead}}, \href{https://doi.org/10.1103/PhysRevLett.119.251304}{\emph{Phys. Rev. Lett.} {\bfseries 119} (2017) 251304} [\href{https://arxiv.org/abs/1710.05901}{{\ttfamily 1710.05901}}].

\bibitem{CreminelliVernizzi2017}
P.~Creminelli and F.~Vernizzi, \emph{{Dark Energy after GW170817 and GRB170817A}}, \href{https://doi.org/10.1103/PhysRevLett.119.251302}{\emph{Phys. Rev. Lett.} {\bfseries 119} (2017) 251302} [\href{https://arxiv.org/abs/1710.05877}{{\ttfamily 1710.05877}}].

\bibitem{Baker2017}
T.~Baker, E.~Bellini, P.~G. Ferreira, M.~Lagos, J.~Noller and I.~Sawicki, \emph{{Strong constraints on cosmological gravity from GW170817 and GRB 170817A}}, \href{https://doi.org/10.1103/PhysRevLett.119.251301}{\emph{Phys. Rev. Lett.} {\bfseries 119} (2017) 251301} [\href{https://arxiv.org/abs/1710.06394}{{\ttfamily 1710.06394}}].

\bibitem{Langlois2018}
D.~Langlois, R.~Saito, D.~Yamauchi and K.~Noui, \emph{{Scalar-tensor theories and modified gravity in the wake of GW170817}}, \href{https://doi.org/10.1103/PhysRevD.97.061501}{\emph{Phys. Rev. D} {\bfseries 97} (2018) 061501} [\href{https://arxiv.org/abs/1711.07403}{{\ttfamily 1711.07403}}].

\bibitem{DESI2025DR2}
{\scshape DESI} collaboration, \emph{{DESI DR2 results. II. Measurements of Baryon Acoustic Oscillations and Cosmological Constraints}}, \href{https://doi.org/10.1103/tr6y-kpc6}{\emph{Phys. Rev. D} {\bfseries 112} (2025) 083515} [\href{https://arxiv.org/abs/2503.14738}{{\ttfamily 2503.14738}}].

\bibitem{Ye2025}
G.~Ye, M.~Martinelli, B.~Hu and A.~Silvestri, \emph{{Hints of Nonminimally Coupled Gravity in DESI 2024 Baryon Acoustic Oscillation Measurements}}, \href{https://doi.org/10.1103/PhysRevLett.134.181002}{\emph{Phys. Rev. Lett.} {\bfseries 134} (2025) 181002} [\href{https://arxiv.org/abs/2407.15832}{{\ttfamily 2407.15832}}].

\bibitem{Babichev2016review}
E.~Babichev, C.~Charmousis and A.~Leh{\'e}bel, \emph{{Black holes and stars in Horndeski theory}}, \href{https://doi.org/10.1088/0264-9381/33/15/154002}{\emph{Class. Quant. Grav.} {\bfseries 33} (2016) 154002} [\href{https://arxiv.org/abs/1604.06402}{{\ttfamily 1604.06402}}].

\bibitem{Babichev2023book}
E.~Babichev, C.~Charmousis and N.~Lecoeur, \emph{{Exact black hole solutions in higher-order scalar-tensor theories}},  in \emph{{Compact Objects in the Universe}}, E.~Papantonopoulos and N.~Mavromatos, eds., (Cham), Springer, (2024), \href{https://arxiv.org/abs/2309.12229}{{\ttfamily 2309.12229}}, \href{https://doi.org/10.1007/978-3-031-55098-0}{DOI}.

\bibitem{BenAchour2025review}
J.~Ben~Achour, \emph{{DHOST theories as disformal gravity: from black holes to radiative spacetimes}}, \href{https://doi.org/10.1140/epjc/s10052-025-13983-8}{\emph{Eur. Phys. J. C} {\bfseries 85} (2025) 424} [\href{https://arxiv.org/abs/2412.04135}{{\ttfamily 2412.04135}}].

\bibitem{Mironov2019wormhole}
S.~Mironov, V.~Rubakov and V.~Volkova, \emph{{More about stable wormholes in beyond Horndeski theory}}, \href{https://doi.org/10.1088/1361-6382/ab2574}{\emph{Class. Quant. Grav.} {\bfseries 36} (2019) } [\href{https://arxiv.org/abs/1812.07022}{{\ttfamily 1812.07022}}].

\bibitem{Mironov2023wormhole}
S.~Mironov, V.~Rubakov and V.~Volkova, \emph{{In hot pursuit of a stable wormhole in beyond Horndeski theory}}, \href{https://doi.org/10.1103/PhysRevD.107.104061}{\emph{Phys. Rev. D} {\bfseries 107} (2023) 104061} [\href{https://arxiv.org/abs/2212.05969}{{\ttfamily 2212.05969}}].

\bibitem{BabichevCharmousis2014}
E.~Babichev and C.~Charmousis, \emph{{Dressing a black hole with a time-dependent Galileon}}, \href{https://doi.org/10.1007/JHEP08(2014)106}{\emph{JHEP} {\bfseries 08} (2014) 106} [\href{https://arxiv.org/abs/1312.3204}{{\ttfamily 1312.3204}}].

\bibitem{Kolevatov2017perturbations}
R.~Kolevatov, S.~Mironov, V.~Rubakov, N.~Sukhov and V.~Volkova, \emph{{Perturbations in generalized Galileon theories}}, \href{https://doi.org/10.1103/PhysRevD.96.125012}{\emph{Phys. Rev. D} {\bfseries 96} (2017) 125012} [\href{https://arxiv.org/abs/1708.04262}{{\ttfamily 1708.04262}}].

\bibitem{MironovShtennikova2023}
S.~Mironov and A.~Shtennikova, \emph{{Perturbations in Horndeski theory above anisotropic cosmological background}}, \href{https://doi.org/10.1134/S002136402360204X}{\emph{JETP Lett.} (2023) } [\href{https://arxiv.org/abs/2305.19171}{{\ttfamily 2305.19171}}].

\bibitem{Kobayashi2012_static}
T.~Kobayashi, H.~Motohashi and T.~Suyama, \emph{{Black hole perturbations in the most general scalar-tensor theory with second-order field equations I: the odd-parity sector}}, \href{https://doi.org/10.1103/PhysRevD.85.084025}{\emph{Phys. Rev. D} {\bfseries 85} (2012) 084025} [\href{https://arxiv.org/abs/1202.4893}{{\ttfamily 1202.4893}}].

\bibitem{Kobayashi2014_static}
T.~Kobayashi, H.~Motohashi and T.~Suyama, \emph{{Black hole perturbation in the most general scalar-tensor theory with second-order field equations II: the even-parity sector}}, \href{https://doi.org/10.1103/PhysRevD.89.084042}{\emph{Phys. Rev. D} {\bfseries 89} (2014) 084042} [\href{https://arxiv.org/abs/1402.6740}{{\ttfamily 1402.6740}}].

\bibitem{Minamitsuji2022b}
M.~Minamitsuji, K.~Takahashi and S.~Tsujikawa, \emph{{Linear stability of black holes with static scalar hair in full Horndeski theories: Generic instabilities and surviving models}}, \href{https://doi.org/10.1103/PhysRevD.106.044003}{\emph{Phys. Rev. D} {\bfseries 106} (2022) 044003} [\href{https://arxiv.org/abs/2204.13837}{{\ttfamily 2204.13837}}].

\bibitem{MironovVolkova2023DPSV}
S.~Mironov and V.~Volkova, \emph{{DPSV trick for spherically symmetric backgrounds}}, \href{https://doi.org/10.1016/j.nuclphysb.2024.116550}{\emph{Nucl. Phys. B} {\bfseries 1004} (2024) 116550} [\href{https://arxiv.org/abs/2306.17791}{{\ttfamily 2306.17791}}].

\bibitem{MironovVolkova2024complete}
S.~Mironov and V.~Volkova, \emph{{Complete stability for spherically symmetric backgrounds in beyond Horndeski theory}}, \href{https://doi.org/10.1142/S0217751X24430115}{\emph{Int. J. Mod. Phys. A} {\bfseries 39} (2024) 2443011} [\href{https://arxiv.org/abs/2404.06297}{{\ttfamily 2404.06297}}].

\bibitem{Boonserm2026}
P.~Boonserm, A.~De~Felice, R.~Nakarachinda, S.~Tsujikawa and P.~Wongjun, \emph{{Perturbations and stability of black holes with static scalar hair in general GLPV theories}},  \href{https://arxiv.org/abs/2609.00664}{{\ttfamily 2609.00664}}.

\bibitem{BoumazaLanglois2025}
H.~Boumaza and D.~Langlois, \emph{{Neutron stars in degenerate higher-order scalar-tensor theories: Axial perturbations}},  \href{https://arxiv.org/abs/2507.02749}{{\ttfamily 2507.02749}}.

\bibitem{Ogawa2016}
H.~Ogawa, T.~Kobayashi and T.~Suyama, \emph{{Instability of hairy black holes in shift-symmetric Horndeski theories}}, \href{https://doi.org/10.1103/PhysRevD.93.064078}{\emph{Phys. Rev. D} {\bfseries 93} (2016) 064078} [\href{https://arxiv.org/abs/1510.07400}{{\ttfamily 1510.07400}}].

\bibitem{TakahashiSuyama2017}
K.~Takahashi and T.~Suyama, \emph{{Linear perturbation analysis of hairy black holes in shift-symmetric Horndeski theories: Odd-parity perturbations}}, \href{https://doi.org/10.1103/PhysRevD.95.024034}{\emph{Phys. Rev. D} {\bfseries 95} (2017) 024034} [\href{https://arxiv.org/abs/1610.00432}{{\ttfamily 1610.00432}}].

\bibitem{TretyakovaTakahashi2017}
D.~A. Tretyakova and K.~Takahashi, \emph{{Stable black holes in shift-symmetric Horndeski theories}}, \href{https://doi.org/10.1088/1361-6382/aa8057}{\emph{Class. Quant. Grav.} {\bfseries 34} (2017) 175007} [\href{https://arxiv.org/abs/1702.03502}{{\ttfamily 1702.03502}}].

\bibitem{Babichev2018PRL}
E.~Babichev, C.~Charmousis, G.~Esposito-Far{\`e}se and A.~Leh{\'e}bel, \emph{{Stability of Black Holes and the Speed of Gravitational Waves within Self-Tuning Cosmological Models}}, \href{https://doi.org/10.1103/PhysRevLett.120.241101}{\emph{Phys. Rev. Lett.} {\bfseries 120} (2018) 241101} [\href{https://arxiv.org/abs/1712.04398}{{\ttfamily 1712.04398}}].

\bibitem{Takahashi2019}
K.~Takahashi, H.~Motohashi and M.~Minamitsuji, \emph{{Linear stability analysis of hairy black holes in quadratic degenerate higher-order scalar-tensor theories: Odd-parity perturbations}}, \href{https://doi.org/10.1103/PhysRevD.100.024041}{\emph{Phys. Rev. D} {\bfseries 100} (2019) 024041} [\href{https://arxiv.org/abs/1904.03554}{{\ttfamily 1904.03554}}].

\bibitem{TomikawaKobayashi2021}
K.~Tomikawa and T.~Kobayashi, \emph{{Perturbations and quasi-normal modes of black holes with time-dependent scalar hair in shift-symmetric scalar-tensor theories}}, \href{https://doi.org/10.1103/PhysRevD.103.084041}{\emph{Phys. Rev. D} {\bfseries 103} (2021) 084041} [\href{https://arxiv.org/abs/2101.03790}{{\ttfamily 2101.03790}}].

\bibitem{Noui2023}
K.~Noui, H.~Roussille and D.~Langlois, \emph{{Axial perturbations of black holes in scalar-tensor gravity: near-horizon behaviour}}, \href{https://doi.org/10.1088/1475-7516/2023/11/040}{\emph{JCAP} {\bfseries 11} (2023) 040} [\href{https://arxiv.org/abs/2304.03055}{{\ttfamily 2304.03055}}].

\bibitem{Charmousis2025axial}
C.~Charmousis, S.~Iteanu, D.~Langlois and K.~Noui, \emph{{Axial perturbations of black holes with primary scalar hair}}, \href{https://doi.org/10.1088/1475-7516/2025/05/102}{\emph{JCAP} {\bfseries 05} (2025) 102} [\href{https://arxiv.org/abs/2503.22348}{{\ttfamily 2503.22348}}].

\bibitem{Babichev2018hamiltonian}
E.~Babichev, C.~Charmousis, G.~Esposito-Far{\`e}se and A.~Leh{\'e}bel, \emph{{Hamiltonian vs stability and application to Horndeski theory}}, \href{https://doi.org/10.1103/PhysRevD.98.104050}{\emph{Phys. Rev. D} {\bfseries 98} (2018) 104050} [\href{https://arxiv.org/abs/1803.11444}{{\ttfamily 1803.11444}}].

\bibitem{deRhamZhang2019}
C.~de~Rham and J.~Zhang, \emph{{Perturbations of Stealth Black Holes in DHOST Theories}}, \href{https://doi.org/10.1103/PhysRevD.100.124023}{\emph{Phys. Rev. D} {\bfseries 100} (2019) 124023} [\href{https://arxiv.org/abs/1907.00699}{{\ttfamily 1907.00699}}].

\bibitem{Langlois2021}
D.~Langlois, K.~Noui and H.~Roussille, \emph{{Black hole perturbations in modified gravity}}, \href{https://doi.org/10.1103/PhysRevD.104.124044}{\emph{Phys. Rev. D} {\bfseries 104} (2021) 124044} [\href{https://arxiv.org/abs/2103.14750}{{\ttfamily 2103.14750}}].

\bibitem{TakahashiMotohashi2021}
K.~Takahashi and H.~Motohashi, \emph{{Black hole perturbations in DHOST theories: master variables, gradient instability, and strong coupling}}, \href{https://doi.org/10.1088/1475-7516/2021/08/013}{\emph{JCAP} {\bfseries 08} (2021) 013} [\href{https://arxiv.org/abs/2106.07128}{{\ttfamily 2106.07128}}].

\bibitem{Minamitsuji2022a}
M.~Minamitsuji, K.~Takahashi and S.~Tsujikawa, \emph{{Linear stability of black holes in shift-symmetric Horndeski theories with a time-independent scalar field}}, \href{https://doi.org/10.1103/PhysRevD.105.104001}{\emph{Phys. Rev. D} {\bfseries 105} (2022) 104001} [\href{https://arxiv.org/abs/2201.09687}{{\ttfamily 2201.09687}}].

\bibitem{Charmousis2026radial}
C.~Charmousis, S.~Iteanu, D.~Langlois and K.~Noui, \emph{{Radial Perturbations of Black Holes in DHOST Theories}},  \href{https://arxiv.org/abs/2606.25972}{{\ttfamily 2606.25972}}.

\bibitem{Libanov2016}
M.~Libanov, S.~Mironov and V.~Rubakov, \emph{{Generalized Galileons: instabilities of bouncing and Genesis cosmologies and modified Genesis}}, \href{https://doi.org/10.1088/1475-7516/2016/08/037}{\emph{JCAP} {\bfseries 08} (2016) 037} [\href{https://arxiv.org/abs/1605.05992}{{\ttfamily 1605.05992}}].

\bibitem{Kobayashi2016nogo}
T.~Kobayashi, \emph{{Generic instabilities of non-singular cosmologies in Horndeski theory: A no-go theorem}}, \href{https://doi.org/10.1103/PhysRevD.94.043511}{\emph{Phys. Rev. D} {\bfseries 94} (2016) 043511} [\href{https://arxiv.org/abs/1606.05831}{{\ttfamily 1606.05831}}].

\bibitem{Rubakov2016wormholes}
V.~A. Rubakov, \emph{{More about wormholes in generalized Galileon theories}}, \href{https://doi.org/10.1134/S0040577916080080}{\emph{Theor. Math. Phys.} {\bfseries 188} (2016) } [\href{https://arxiv.org/abs/1601.06566}{{\ttfamily 1601.06566}}].

\bibitem{EvseevMelichev2018}
O.~A. Evseev and O.~I. Melichev, \emph{{No static spherically symmetric wormholes in Horndeski theory}}, \href{https://doi.org/10.1103/PhysRevD.97.124040}{\emph{Phys. Rev. D} {\bfseries 97} (2018) 124040} [\href{https://arxiv.org/abs/1711.04152}{{\ttfamily 1711.04152}}].

\bibitem{oddSector}
S.~Mironov, M.~Sharov and V.~Volkova, \emph{{Linear stability of a time-dependent, spherically symmetric background in beyond Horndeski theory and the speed of gravity waves}}, \href{https://doi.org/10.1140/epjc/s10052-024-13730-5}{\emph{Eur. Phys. J. C} {\bfseries 85} (2025) 50} [\href{https://arxiv.org/abs/2408.01480}{{\ttfamily 2408.01480}}].

\bibitem{MironovSharov2025}
S.~Mironov and M.~Sharov, \emph{{No-Go theorem in the cubic subclass of Horndeski theory for spherically symmetric dynamical background}}, \href{https://doi.org/10.1103/PhysRevD.111.084033}{\emph{Phys. Rev. D} {\bfseries 111} (2025) 084033} [\href{https://arxiv.org/abs/2411.05416}{{\ttfamily 2411.05416}}].

\bibitem{Tsujikawa2022}
S.~Tsujikawa, \emph{{Instability of hairy black holes in regularized 4-dimensional Einstein-Gauss-Bonnet gravity}}, \href{https://doi.org/10.1016/j.physletb.2022.137329}{\emph{Phys. Lett. B} (2022) 137329} [\href{https://arxiv.org/abs/2205.09932}{{\ttfamily 2205.09932}}].

\bibitem{Minamitsuji2024}
M.~Minamitsuji, S.~Mukohyama and S.~Tsujikawa, \emph{{Angular and radial stabilities of spontaneously scalarized black holes in the presence of scalar-Gauss-Bonnet couplings}}, \href{https://doi.org/10.1103/PhysRevD.109.104057}{\emph{Phys. Rev. D} {\bfseries 109} (2024) 104057} [\href{https://arxiv.org/abs/2403.10048}{{\ttfamily 2403.10048}}].

\bibitem{ReggeWheeler1957}
T.~Regge and J.~A. Wheeler, \emph{{Stability of a Schwarzschild singularity}}, \href{https://doi.org/10.1103/PhysRev.108.1063}{\emph{Phys. Rev.} {\bfseries 108} (1957) 1063}.

\end{thebibliography}\endgroup

\end{document}